\documentclass[useAMS,usegraphicx]{mn2e}

\def\H2{H$_2$}
\def\nH{n_{\rm H}}

\def\fH2{f_{\rm H_2}}
\def\zvir{z_{\rm vir}}
\def\Mvir{M_{\rm vir}}
\def\apj{ApJ}
\def\apjs{ApJS}
\def\aap{A\&A}
\def\mnras{MNRAS}
\def\pasj{PASJ}
\def\aj{AJ}

\begin{document}
\title[Dust and early galaxy evolution]{Effects of dust grains
on early galaxy evolution}
\author[H. Hirashita \& A. Ferrara]{Hiroyuki
Hirashita$^{1,2}$ and Andrea Ferrara$^{1}$\\
$^1$ Osservatorio Astrofisico di Arcetri, Largo Enrico Fermi, 5,
     50125 Firenze, Italy \\
$^2$ Postdoctoral Fellow of the Japan Society for the Promotion of
     Science (JSPS) for Research Abroad}
\date{Accepted 2002 August 13}
\pubyear{2002} \volume{000} \pagerange{1}
\twocolumn

\maketitle \label{firstpage}
\begin{abstract}
Stars form out of molecular gas and supply dust grains
during their last evolutionary stages; in turn hydrogen
molecules (\H2) are produced more efficiently on dust grains.
Therefore, dust can drastically accelerate \H2 formation, leading
to an enhancement of star formation activity. In order to examine
the first formation of stars and dust in
galaxies, we model the evolution of galaxies in the redshift range
of $5<z<20$. In particular, we focus on the interplay
between dust formation in Type II supernova ejecta and H$_2$
production on dust grains. Such effect causes an enhancement of
star formation rate by an order of magnitude on a timescale
($\sim 3$--5 galactic dynamical times) shorter than the Hubble
timescale. We also find that about half of the radiative energy
from stars is reprocessed by dust grains and is finally radiated
in the far infrared (FIR). For example, the typical star
formation rate, FIR and ultraviolet (UV) luminosity of a galaxy with
$\Mvir =10^{11.5}~M_\odot$ (virial mass) and $\zvir =5$
(formation redshift) are 3 $M_\odot$ yr$^{-1}$,
$4\times 10^9~L_\odot$, and $3\times 10^9~L_\odot$, respectively.
This object will be detected by both ALMA and {\it NGST}.
Typical star formation rates and
luminosities (FIR, UV and metal-line luminosities) are calculated
for a large set of $(\Mvir ,\,\zvir )$. Using these results and
the Press-Schechter formalism, we
calculate galaxy number counts and integrated light from
high-redshift ($z>5$) galaxies in sub-millimetre and
near-infrared bands. We find that: i) ALMA can detect dust
emission from $\mbox{several}\times 10^3$ galaxies per square
degree, and ii) {\it NGST} can detect the stellar emission from
$10^6$ galaxies per square degree. Further observational checks
of our
predictions include the integrated flux of metal (oxygen and
carbon) lines;
these lines can be used to trace the chemical enrichment and the
gas density in early galactic environments. We finally discuss
possible color selection strategies for high-redshift galaxy
searches.
\end{abstract}

\section{Introduction}

In order to understand the chemical and thermodynamical state
of the interstellar medium (ISM) of primeval galaxies, dust
formation needs to be considered. Even in metal poor galaxies, dust
grains can drastically accelerate the formation rate of
molecular hydrogen (H$_2$), expected to be the most abundant molecule
in the ISM (Hirashita, Hunt, \& Ferrara 2002a).
Hydrogen molecules emit vibrational-rotational lines,
thus cooling the gas. This process is particularly important
to understand the formation of stars in metal-poor primeval
galaxies (e.g., Matsuda, Sato, \& Takeda 1969; Omukai \& Nishi 1998;
Nishi \& Susa 1999; Bromm, Coppi, \&
Larson 2002; Abel, Bryan, \& Norman 2002; Nakamura \& Umemura 2002;
Kamaya \& Silk 2002; Ripamonti et al.\ 2002).
The important role of dust on the enhancement of \H2
abundance is also suggested by observations of damped Ly$\alpha$
systems (DLAs; Ge, Bechtold,
\& Kulkarni 2001; cf.\ Petitjean, Srianand, \& Ledoux 2000).

The existence of dust in young galaxies is naturally expected
because Type II supernovae (SNe II) are shown to produce dust
grains  (e.g.,
Dwek et al.\ 1983; Moseley et al.\ 1989;
Kozasa, Hasegawa, \& Nomoto 1991;
Todini \& Ferrara 2001). Since the lifetime of
SN II progenitors (massive stars) is short, SNe II are the
dominant production source for of dust grains in young
($<1$ Gyr) star-forming galaxies. The winds of evolved low-mass
stars contribute to dust formation considerably in nearby galaxies
(Gehrz 1989), but the cosmic time is not long enough for such stars
to evolve at high redshift ($z>5$), when all galaxies should have
ages smaller than $\sim 1$ Gyr. However,
dust is also destroyed by SN shocks
(McKee 1989; Jones, Tielens, \& Hollenbach 1996). The detailed 
modelling of dust
evolution in galaxies therefore requires an accurate treatment of
both types of processes (for recent modeling, see
e.g., Edmunds 2001; Hirashita, Tajiri, \& Kamaya 2002b).

Here we model the evolution of dust content in primeval galaxies.
We adopt the results of Todini \& Ferrara (2001) for the dust
formation rate in SNe II. Although further discussion on their
application of nucleation theory is necessary (e.g.,
Frenklach \& Feigelson 1997), their results have been
successfully applied to the interpretation not only of the
properties of SN 1987A but also of the FIR properties of the
young dwarf galaxy SBS 0335$-$052 (Hirashita et al.\ 2002a).

One of the most direct observational constraints for the
evolution of dust content in galaxies comes from the
far-infrared (FIR) properties of galaxies. Dust grains absorb
stellar light and reemit it in FIR.
Recent observations by the Submillimetre
Common-User Bolometer Array (SCUBA) and the {\it Infrared Space
Observatory} ({\it ISO}) have made it possible to study
galaxy evolution in the FIR band up to $z\la 3$
(Smail et al.\ 1998). The detection of the cosmic
infrared--submillimetre (sub-mm) background by the
{\it COsmic Background Explorer} ({\it COBE}\,)
(Puget et al.\ 1996; Fixsen et al.\ 1998) has also provided
crucial information
on the star formation history of galaxies in the universe
(e.g., Dwek et al.\ 1998). Some theoretical works have
modelled the FIR evolution of galaxies up to $z\sim 5$
(Tan, Silk, \& Balland 1999; Pei, Fall, \& Hauser 1999;
Takeuchi et al.\ 2001a; Xu et al.\ 2001;
Pearson 2001; Totani \& Takeuchi 2002), and the FIR luminosity
of galaxies per unit comoving volume seems to be much higher
at $z\sim 1$ than at $z\sim 0$ (see also Elbaz et al.\ 2002).
However, such a strong ``evolution'' beyond $z=2$ has been
excluded (Gispert, Lagache, \& Puget 2000; Malkan \& Stecker 2001;
Takeuchi et al.\ 2001a).

Although there is clear evidence for the existence of dust
in galaxies at $z\la 5$ (Armus et al.\ 1998;
Soifer et al.\ 1998), few works focusing on the early dust
formation in galaxies
exist. Some ``semi-analytic'' works have included the
dust formation in the early galaxy evolution (e.g.,
Devriendt \& Guiderdoni 2000; Granato et al.\ 2000), but there
has been no study treating the dust
formation, the molecular formation on grain surfaces, and the
star formation history in a consistent manner. Therefore,
in this paper, we model the three processes consistently so that
we can obtain an observational strategy under a consistent
scenario for the early evolution of galaxies.

In order to understand which physical processes govern dust
formation, observations at sub-mm
wavelengths ($300~\mu{\rm m}\la\lambda\la 1$ mm) are crucial.
For high-redshift objects, redshifted FIR radiation,
i.e., sub-mm light, should be observed to detect the dust
emission. In particular, detecting the sub-mm radiation from
galaxies at $z>5$ requires
more sensitive and high-resolution observations
(e.g., Takeuchi et al.\ 2001b). A future ground-based
interferometric facility, the Atacama Large Millimeter Array
(ALMA\footnote{http://www.eso.org/projects/alma/}), can be used
to study such high-redshift galaxies. The detected amount of metals
and stars can be used to constrain the
galaxy evolution through a chemical evolution model
(Tinsley 1980). Redshifted sub-mm metal emission lines can also be
observed with ALMA. This can directly constrain the abundance of
metals formed in the early epoch of galaxy evolution.
(Oh et al.\ 2002 have also proposed to probe high-redshift
intergalactic medium metallicity by metal absorption lines.)
In order to detect the stellar light from the
high-redshift universe, observations by the {\it Next Generation
Space Telescope} ({\it NGST}~\footnote{http://ngst.gsfc.nasa.gov/})
in near infrared (NIR) will be particularly suitable. At 2 $\mu$m,
for example, we can observe the
$\sim 2000$ \AA\ ultraviolet (UV) light radiated from a galaxy at
$z\sim 10$. Therefore, our scenario will become testable in the near
future. This means that it is worth constructing a consistent model
for the high-redshift galaxy evolution.

For any observational facility, statistical properties of galaxies
should be discussed to obtain a general picture of galaxy evolution.
Two quantities are
particularly important for statistical purposes: galaxy number
counts (the number of galaxies as a function of observed flux) and
integrated light (the sum of
the flux from all the galaxies considered; Hauser \& Dwek 2001 for
a review). In this paper, therefore, we estimate the contribution
of high-redshift galaxies to these two quantities.

Throughout this paper, we assume a flat cold dark matter (CDM)
cosmology with a cosmological constant. The values of quantities
are the same as those in Mo \& White (2002)
($\Omega_{\rm M}=0.3$, $\Omega_\Lambda =0.7$, and
$H_0\equiv 100 h$ km s$^{-1}$ Mpc$^{-1}=70$ km s$^{-1}$ Mpc$^{-1}$).
The baryon density parameter is assumed to be
$\Omega_{\rm b}=0.02h^{-2}$. For the power spectrum of the density
fluctuation, $n=1$ and $\sigma_8=0.9$ are
adopted. We first model the physical state of gas and the content
of dust and metals in a galaxy during its early evolutionary stage
(\S~\ref{sec:each}). There we also model the luminosities of FIR,
UV, and metal lines. The result of our model for fiducial galaxies
are presented in \S~\ref{sec:result}. Based on these results, we
next calculate the galaxy number counts and the
integrated light at various wavelengths in \S~\ref{sec:nc}.
We discuss the observational implications of our results in
\S~\ref{sec:discussion}.

\section{Galaxy evolution model}\label{sec:each}

We aim at modelling the star formation rate during the early
stage of galaxy evolution. We consider that stars form only in
molecular environments, where cooling to a low temperature is
possible. We solve the evolution of molecular content in a way
consistent with the star formation history and the dust supply
from stars. Some of the past studies (e.g.,
Norman \& Spaans 1996) have also considered similar processes,
but we now tie star formation to the molecular content
in determining the star formation rate (\S~\ref{subsec:sfr}).

Before $z=5$, all the galaxies are younger than the cosmic
age\footnote{The cosmic ages for the adopted cosmology at $z=5$,
10, and 20 are 1.2, 0.47, and 0.18 Gyr, respectively.} of 1.2 Gyr.
Although dust can play an
important role in forming hydrogen molecules, the model for the
evolution of dust content in the early epoch of the universe has
not been
developed yet. Since planned facilities will enable us to observe
dust emission from $z>5$, it is worth constructing a theoretical
framework for the dust evolution at $z>5$.
For observational strategies, we estimate the radiative energy
at various wavelengths. For $z>20$, the number of galaxies
which survive stellar feedback is negligible
(Ciardi et al.\ 2000). Therefore, we concentrate ourselves on
galaxies formed between $z=20$ and 5.

\subsection{Interstellar medium evolution}
\label{subsec:phys_state}

\subsubsection{Gas density}\label{subsubsec:density}

The star formation process is affected by the physical state of
the ISM. In particular, cooling by molecular hydrogen plays an
important role in star formation. Although the law governing
the galactic star formation rate is not established at all for
primeval galaxies (Kennicutt's
empirical law is derived from observations of nearby galaxies;
Kennicutt 1998), the abundance of H$_2$ should be ultimately a
key parameter as stars are only seen to form in molecular
complexes (see e.g., Wilson et al.\ 2000 for a recent observation
of a nearby galaxy).
Moreover, enhancement of molecular gas formation
is shown to result in active star formation (for recent results,
see e.g., Walter et al.\ 2002).
We define the
molecular fraction of hydrogen, $\fH2$, as
\begin{eqnarray}
\fH2\equiv 2n_{\rm H_2}/\nH\, ,\label{eq:mol_frac}
\end{eqnarray}
where $n_{\rm H_2}$ and $\nH$ are the number densities of
hydrogen nuclei and hydrogen molecules, respectively
(i.e., if all the hydrogen nuclei are in the molecular form,
$\fH2 =1$). The importance of $\fH2$ in the star formation law is
modeled in \S~\ref{subsec:sfr}.

We consider a pure hydrogen gas, and neglect helium in this paper
for the following two reasons. First, helium
has little influence on the electron abundance
(i.e., the formation rate of \H2 in the gas phase is not affected
by the presence of
He; Hutchings et al.\ 2002). Second, the existence of helium does
not affect the temperature of gas and thus it does not change
the reaction rates concerning \H2 formation
Kitayama et al.\ (2000). In order to examine the time evolution
of $\fH2$, we solve a set of reaction equations describing the
\H2 formation and destruction.

First, we must estimate the number density of the hydrogen gas
$\nH$, because it affects both the reaction rate and the cooling
rate. The objects that can form stars continuously against the
heating from interstellar UV (IUV)\footnote{In order to avoid
confusion between external
(background) and internal (interstellar) UV fields, we call the
UV from stars within the galaxy ``IUV''.}
radiation and stellar kinetic energy input have virial
temperature typically larger than $10^4$ K
(Ciardi et al.\ 2000). Since we are interested in the star
formation activity of galaxies, we only consider objects whose
virial temperature is larger than such value. If the gas
temperature is larger than $10^4$ K, cooling by hydrogen atomic
lines is efficient. Then the gas loses pressure
and collapses in the gravitational field of dark matter. Even in
the presence of UV background radiation, objects with such a
high virial temperature can collapse because of efficient cooling
and \H2 self-shielding (Kitayama et al.\ 2001).
The collapse increases the \H2 column density further, and
as a result, external UV shielded by \H2 has
little effect on the
temperature and dynamics of the gas after the collapse.
Therefore, we neglect the external UV radiation field.

Since the gas cooling time is much shorter than the Hubble
timescale for the objects of interest in this paper
(Madau, Ferrara, \& Rees 2001), we expect that a significant
fraction of baryons finally collapses in the dark matter
potential. If the halo is rotating, the gas will collapse in a
centrifugally supported disk. The radius ($r_{\rm disk}$) and
the scale height
($H$) of the disk are determined following
Ferrara, Pettini, \& Shchekinov (2000) (see also
Norman \& Spaans 1996; Ciardi \& Loeb 2000). By considering the
conservation of angular momentum and assuming a typical value
for the spin parameter
($\lambda =0.04$; Barnes \& Efstathiou 1987;
Steinmetz \& Bartelmann 1995), we obtain
$r_{\rm disk}\simeq 0.18r_{\rm vir}$, and the radius of
the dark halo, $r_{\rm vir}$,
is estimated in terms of the mass of the dark halo, $\Mvir$,
and the redshift of virialisation, $\zvir$, as
\begin{eqnarray}
\frac{4\pi}{3}r_{\rm vir}^3\Delta_{\rm c}\,\rho_{\rm c0}
\Omega_{\rm M}(1+\zvir )^3=\Mvir\, ,
\end{eqnarray}
where $\rho_{\rm c0}\equiv 3c^2H_0^2/8\pi G$ and
$\Delta_{\rm c}$ are the critical density of the universe at
$z=0$ and the overdensity of an object formed at $\zvir$,
respectively. The spherical collapse model (e.g., Peebles 1980)
predicts $\Delta_{\rm c}(\zvir)\simeq 180$. We adopt the
fitting formula
given by Kitayama \& Suto (1996) for $\Delta_{\rm c}(\zvir)$
as
\begin{eqnarray}
\Delta_{\rm c}(\zvir )\simeq 18\pi^2[1+0.4093
(1/\Omega_{\rm v}-1)^{0.9052}]\, ,
\end{eqnarray}
where $\Omega_{\rm v}$ is the density parameter at $\zvir$
given by
\begin{eqnarray}
\Omega_{\rm v}=
\frac{\Omega_{\rm M}(1+\zvir )^3}
{\Omega_{\rm M}(1+\zvir )^3+(1-\Omega_{\rm M}-\Omega_\Lambda)
(1+\zvir )^2+\Omega_\Lambda}\, .
\end{eqnarray}
This formula is applicable only to a flat universe
($\Omega_{\rm M}+\Omega_\Lambda =1$) with $\Omega_{\rm M}<1$.

Using $r_{\rm disk}$ and $H$ (the scale height of disk),
$\nH$ is estimated from
\begin{eqnarray}
\nH\simeq
\frac{\Mvir\Omega_{\rm b}}{\pi r_{\rm disk}^22Hm_{\rm H}\Omega_{\rm M}}\,
,
\end{eqnarray}
where $m_{\rm H}$ is the mass of a hydrogen atom.
We have assumed that the mass of
a galaxy is dominated by the dark halo and that the mass of
the gas is $\Mvir\Omega_{\rm b}/\Omega_{\rm M}$.

An object collapsed at $\zvir$ with a mass of $\Mvir$ has a
virial temperature, $T_{\rm vir}$, defined as
\begin{eqnarray}
T_{\rm vir}\equiv\frac{G\mu \Mvir}{3k_{\rm B}r_{\rm vir}}\, ,
\label{eq:T_vir}
\end{eqnarray}
where $G$ is the gravitational constant, $k_{\rm B}$ is the
Boltzmann constant, and $\mu$ is the mean molecular weight. The
factor ``3'' can be different depending on the radial
profile of gas. Therefore, the definition of $T_{\rm vir}$
contains an uncertainty of the order unity. Since we are
considering a pure hydrogen gas,
$\mu =(1+x)^{-1}m_{\rm H}$, where $x$
is the ionisation degree (fraction of hydrogen nuclei in the
ionized state) which is derived from equation
(\ref{eq:ion_deg}) below. The circular velocity, $v_{\rm c}$,
is defined as
\begin{eqnarray}
v_{\rm c}\equiv\sqrt\frac{G\Mvir}{r_{\rm vir}}\, ;
\end{eqnarray}
we also define the circular timescale, $t_{\rm cir}$, as
\begin{eqnarray}
t_{\rm cir}\equiv\frac{2\pi r_{\rm disk}}{v_{\rm c}}\, .
\end{eqnarray}

The disk thickness relative to the radius is estimated by using
the formalism in Ferrara et al.\ (2000). Using $T_{\rm vir}$
(eq.\ \ref{eq:T_vir}) for the gas temperature,
$H/r_{\rm disk}$ is estimated to be $\sim 0.1$. We fix this
thickness for the disk throughout the time evolution of
each galaxy. Probably $H$ could decrease further as gas cools,
but it is
difficult to include a dynamical evolution of gas into
our one-zone model. Moreover, turbulent
energy supplied by SNe II increases $H$, an effect which is
hard to quantify in the framework of this paper. The
resulting gas density roughly follows the scaling relation:
$\nH\sim 80[(1+\zvir)/10]^3~{\rm cm}^{-3}$. This is
higher than the typical density of the local interstellar
medium ($\nH\sim 1~{\rm cm}^{-3}$), reflecting the high-density
environment of high-redshift universe. However, that density
is lower than that calculated by Norman \& Spaans (1996)
because of our higher $H/r_{\rm disk}$. Even in our case,
the gas cools to reach 300 K on a short ($\sim t_{\rm cir}$)
timescale and star formation becomes possible anyway. The
detailed treatment of $H/r_{\rm disk}$ does not affect any of
the following conclusions except for the metal-line emission,
for which we will discuss the uncertainty
caused by this factor (\S~\ref{subsubsec:metal}).
The gas
consumption into stars is also neglected. This is a valid
assumption in this paper, because in
our calculation star formation history is only traced until
$\la 30$\% of the gas content is consumed.

\subsubsection{Chemistry and cooling}\label{subsubsec:chemi}

We calculate the time evolution of ionisation degree ($x$),
molecular fraction ($f_{\rm H_2}$; equation \ref{eq:mol_frac}) and
gas temperature ($T$) of the hydrogen gas. The molecular fraction
is particularly relevant here, because it determines the final
cooling rate of gas when stars form
($T\sim 300$ K). In order to determine $f_{\rm H_2}$, we also
calculate $T$ and $x$. The equations are basically the same as
those of Hutchings et al.\ (2002) except for the
photo-processes (reaction of gas with photons from stars) and
the formation of H$_2$ on dust grains. We summarise the
reactions considered in this paper and their
rate coefficients ($R_n$; $n=0$, ..., 11) in
Table \ref{tab:reaction}. In Table \ref{tab:reaction}, we also list
the reaction coefficient for the \H2 formation on dust
($R_{\rm dust}$; see Appendix \ref{app:surface_rate} for details;
see also Haiman, Rees, \& Loeb 1996 for the list of important
reactions). In Table \ref{tab:photo}, we list the photo-processes
and their
cross sections. The reaction rates of those photo-processes are
expressed as $\Gamma_n$ ($n=12$, ..., 15).

\begin{table*}
\begin{center}
\caption{Reaction rates needed to calculate the abundance of H$_2$.
For the references, see 1) Omukai (2000); 2) Galli \& Palla (1998);
3) this paper. The unit of the gas temperature $T$ is K unless
otherwise stated}
\begin{tabular}{llll}\hline
No. & Reaction & Rate [cm$^3$ s$^{-1}$] & Ref. \\ \hline
& & &  \\
1 & ${\rm H+e^{-}\longrightarrow H^++2e^-}$ &
  $\exp [-32.71+13.54\ln (T({\rm eV}))-5.379(\ln (T({\rm eV})))^2+
  1.563(\ln (T({\rm eV})))^3$ & 1 \\
  & & $~-0.2877(\ln (T({\rm eV})))^4+3.483\times
  10^{-2}(\ln (T({\rm eV})))^5-2.632\times 10^{-3}
  (\ln (T({\rm eV})))^6$ & \\
  & & $~+1.120\times 10^{-4}(\ln (T({\rm eV})))^7-2.039\times
  10^{-6}(\ln (T({\rm eV})))^8]$ & \\
2 & ${\rm H^++e^-\longrightarrow H+\gamma}$ &
  $\exp [-28.61-0.7241(\ln (T({\rm eV})))-2.026\times 10^{-2}(\ln
  (T({\rm eV})))^2$ & 1 \\
  & & $~-2.381\times 10^{-3}(\ln (T({\rm eV})))^3-3.213\times
  10^{-4}(\ln (T({\rm eV})))^4-1.422\times 10^{-5}
  (\ln (T({\rm eV})))^5$ & \\
  & & $~+4.989\times 10^{-6}(\ln (T({\rm eV})))^6
  +5.756\times 10^{-7}(\ln (T({\rm eV})))^7-1.857\times 10^{-8}
  (\ln (T({\rm eV})))^8$ & \\
  & & $~-3.071\times 10^{-9}(\ln (T({\rm eV})))^9]$ & \\
3 & ${\rm H+e^{-}\longrightarrow H^-+\gamma}$ &
  $1.4\times 10^{-18}T^{0.928}\exp (-T/1.62\times 10^4)$ & 1 \\ 
4 & ${\rm H^-+H\longrightarrow H_2+e^-}$ &
  $4.0\times 10^{-9}T^{-0.17}~(T>300)$; $1.5\times 10^{-9}~(T<300)$ & 1 \\
5 & ${\rm H^-+H^+\longrightarrow 2H}$ &
  $5.7\times 10^{-6}T^{-1/2}+6.3\times 10^{-8}-9.2\times 10^{-11}T^{1/2}
  +4.4\times 10^{-13}T$ & 1 \\
6 & ${\rm H+H^+\longrightarrow H_2^++\gamma}$ &
  ${\rm dex}[-19.38-1.523\log_{10}T+1.118(\log_{10}T)^2-0.1269(
  \log_{10}T)^3]$ & 1 \\
7 & ${\rm H_2^++H\longrightarrow H_2+H^+}$ & $6.4\times 10^{-10}$ & 1 \\
8 & ${\rm H_2^++e^-\longrightarrow 2H}$ & $2.0\times 10^{-7}T^{-1/2}$ & 1 \\
9 & ${\rm H_2+H^+\longrightarrow H_2^++H}$ &
  $3.0\times 10^{-10}\exp (-21050/T)$ $(T<10^4)$ & 2 \\
  & & $1.5\times 10^{-10}\exp (-14000/T)$ $(T>10^4)$ & \\
10 & ${\rm H_2+H\longrightarrow 3H}$ & $k_{\rm H}^{1-a}k_{\rm L}^a$
  & 1 \\
  & & ~~$k_{\rm L}=1.12\times 10^{-10}\exp (-7.035\times 10^4/T)$ & \\
  & & ~~$k_{\rm H}=6.5\times 10^{-7}T^{-1/2}\exp (-5.2\times 10^4/T)[
  1-\exp (-6000/T)]$ & \\
  & & ~~$a=4.0-0.416\log_{10}(T/10^4)-0.327(\log_{10}(T/10^4))^2$ & \\
11 & ${\rm H_2+e^-\longrightarrow 2H+e^-}$ &
  $4.4\times 10^{-10}T^{0.35}\exp (-1.02\times 10^5/T)$
  & 1 \\
dust & ${\rm H+H+\mbox{grain}\longrightarrow H_2+\mbox{grain}}$ &
  $2.8\times 10^{-15}(T/100~{\rm K})^{1/2}$  if $T<300$ & 3 \\
  & & 0 if $T>300$ see
  Appendix \ref{app:surface_rate} &  \\
\hline
\end{tabular}
\label{tab:reaction}
\end{center}
\end{table*}

\begin{table*}
\begin{center}
\caption{Cross sections for photo-ionisation and photo-dissociation
processes. The unit of $\nu$ (frequency of light) is Hz.
For the references, see 1) Kitayama \& Ikeuchi (2000);
2) Abel et al.\ (1997); 3) Tegmark et al.\ (1997)}
\begin{tabular}{lllll}\hline
No. & Reaction & cross section & $\nu$ range & Ref. \\
& & (cm$^2$) & (Hz) & \\ \hline
12 & ${\rm H+\gamma\longrightarrow H^++e^-}$ &
  $6.30\times 10^{-18}(\nu /3.3\times 10^{15})^{-3.0}$ &
  $\nu >3.3\times 10^{15}$ & 1 \\
   & & effect of optical depth also included in the text & & \\
13 & ${\rm H_2+\gamma\longrightarrow H_2^*\longrightarrow 2H}$ &
  see equation (\ref{eq:H2_diss}) & & 2 \\
14 & ${\rm H^-+\gamma\longrightarrow H+e^-}$ &
  $3.486\times 10^{-16}(x-1)^{3/2}/x^{3.11}$
  ($x\equiv \nu /1.8\times 10^{14}$) & $\nu >1.8\times 10^{14}$ & 3 \\
15 & ${\rm H_2^++\gamma\longrightarrow H+H^+}$ &
  $7.401\times 10^{-18}\,{\rm dex}(-x^2-0.0302x^3-0.0158x^4)$
  & $\nu >6.4\times 10^{14}$ & 3 \\
  & & ($x\equiv 2.762\ln (\nu /2.7\times 10^{15}$) & & \\
\hline
\end{tabular}
\label{tab:photo}
\end{center}
\end{table*}

The time evolution of the ionizing degree is described as
\begin{eqnarray}
\frac{dx}{dt}  = xf_0R_1\nH -x^2R_2\nH +  \Gamma_{12}f_0\, ,
\label{eq:ion_deg}
\end{eqnarray}
where $f_{\rm 0}\equiv 1-x-\fH2$ is the neutral
fraction of hydrogen. The terms on the right-hand side are the
rates of collisional ionisation, recombination, and
photo-ionisation. Next the time evolution of the molecular
fraction is written as
\begin{eqnarray}
\frac{df_{\rm H_2}}{dt} & = & 2f_0^2x\nH (R_{\rm eff,1}
+R_{\rm eff,2})+2R_{\rm dust}{\cal D}\nH f_0\nonumber\\
& - & \fH2\nH (x^2R_{\rm eff,3}+f_0R_{10}+xR_{11})-\Gamma_{13}
\fH2\, ,\label{eq:H2_form_rate}
\end{eqnarray}
where
\begin{eqnarray}
R_{\rm eff,1}\equiv\frac{R_3R_4}{f_0R_4+xR_5+\Gamma_{14}/\nH}
\, ,
\end{eqnarray}
\begin{eqnarray}
R_{\rm eff,2}\equiv
\frac{R_6R_7}{f_0R_7+xR_8+\Gamma_{15}/\nH}\, ,
\end{eqnarray}
are the effective formation rates of \H2 including the effect of
destruction rate of H$^-$ and H$_2^+$, respectively, and
\begin{eqnarray}
R_{\rm eff,3}\equiv\frac{R_9R_8}{f_0R_7+xR_8+\Gamma_{15}/\nH}\, ,
\end{eqnarray}
is the destruction of H$_2^+$ due to H$^-$ collisions. On the
right-hand side in equation (\ref{eq:H2_form_rate}), we estimate
the rates of the formation in the gas phase, destruction
via reactions 9--11 of Table \ref{tab:reaction},
photo-dissociation, and formation on grains (${\cal D}$ is the
dust-to-gas mass ratio).
We also list the \H2 formation rate on grains, $R_{\rm dust}$ in
Table \ref{tab:reaction} (the details are given in
Appendix \ref{app:surface_rate}). The treatment of
photo-ionisation and \H2 photo-dissociation rates is described
in the next paragraph.
For the initial conditions on $x$ and $\fH2$ at $t=0$
($z=\zvir$), we assume the equilibrium values determined by
$T_{\rm vir}$ and $\nH$. The result is, however, insensitive to
the choice of these values, and as gas cools the molecular
fraction always reaches $\fH2\sim\mbox{a few}\times 10^{-3}$
by $t\sim 10^7$ yr, when the
electron abundance becomes too small to produce further \H2 in the
gas phase.

Equation (A20) of Kitayama \& Ikeuchi (2000) gives $\Gamma_{\rm 12}$
($\Gamma_{\rm HI}$ in their notation) as a function of the incident
UV intensity and the H {\sc i} column density. Although derived for
background UV radiation, their formula is applicable to the
IUV field as well. We assume that the photon paths are optically
thin against ionisation of
H$^-$ and dissociation of H$_2^+$. We also use their
equation (A21) to estimate the photo-ionisation heating. The effect
of radiative transfer is included for photons ionizing
H in the form of the column density of
H {\sc i}, $N_{\rm HI}$. We estimate the column density by
$N_{\rm HI}\simeq \nH (1-x-\fH2 )r_{\rm disk}$. In order to
use the formulation by Kitayama \& Ikeuchi (2000), we describe the
spectrum of the incident IUV radiation from stars by a power law
with an index $\alpha$:
\begin{eqnarray}
I_{\rm IUV}(\nu )=I_0(\nu_{\rm HI})\left(\frac{\nu}{\nu_{\rm HI}}
\right)^{-\alpha}\, ,
\end{eqnarray}
where $\nu$ is the frequency of
photons, $I_0(\nu_{\rm HI})$
is the intensity at the ionisation frequency of neutral
hydrogen ($\nu_{\rm HI}=3.3\times 10^{15}$ Hz).
We use the same spectrum to estimate $\Gamma_n$ ($n=13$, 14,
15) by extending the spectrum down to
$1.8\times 10^{14}~{\rm Hz}$ corresponding to the photon
energy ($h\nu$) of 0.74 eV. This corresponds to the threshold
energy for the photo-ionisation of H$^-$ (Table
\ref{tab:photo}). $I_{\rm IUV}(\nu )$ could be obtained from
a synthetic spectrum of stellar populations, but in this
paper we simply set
$\alpha=5$ for the following two reasons: i) The spectral
shape is uncertain because of the interstellar dust extinction.
In particular, little is known about extinction curve in an
extremely metal-poor environment; ii) Even in the case of
$\alpha=5$, where the largest number of \H2 dissociating photons
are produced among the four spectra examined in
Kitayama et al.\ (2001), dissociation of H$_2$ is negligible in
the presence of
dust grains. The normalisation of the intensity is determined
from
\begin{eqnarray}
\frac{L_{\rm UV,0}\exp (-\tau_{\rm disk})}{4\pi r_{\rm disk}^2}=
\int_{\nu_{\rm min}}^{\infty}
I_{\rm IUV}(\nu )\, d\nu\,,\label{eq:UVnormalize}
\end{eqnarray}
where we define $\nu_{\rm min}$ as the minimum frequency where
OB stars dominate the radiative energy of star-forming galaxies
($\simeq 10^{15}$ Hz),
$L_{\rm UV,0}$ is the UV luminosity estimated in equation
(\ref{eq:uv_lum}), and $\tau_{\rm disk}$ is the typical dust optical
depth in the disk. This typical optical depth can be simply
estimated by multiplying the typical column density of dust
in the direction of disk plane, $n_{\rm dust}r_{\rm dust}$, by
the absorption cross section of dust against UV light,
$\pi a^2Q_{\rm UV}$, as
\begin{eqnarray}
\tau_{\rm disk}=\pi a^2n_{\rm dust}r_{\rm disk}\, .
\end{eqnarray}
where $a$ is the grain radius (spherical grains are assumed).
The method for estimating $L_{\rm UV,0}$ and $n_{\rm dust}$
will be described later in \S~\ref{subsubsec:UV_FIR}.

The \H2 photo-dissociation cross section is estimated
from the rate given by Abel et al.\ (1997). However, if the column
density of
H$_2$ becomes larger than $10^{14}$ cm$^{-2}$, self-shielding
effects become important (Draine \& Bertoldi 1996).
Therefore, we use the following expression for the
H$_2$ dissociation rate:
\begin{eqnarray}
\Gamma_{13} & = & (4\pi )\, 1.1\times 10^8I_{\rm IUV}
(3.1\times 10^{15}~{\rm Hz})\nonumber \\
& \times & \left(
\frac{\nH\fH2 r_{\rm disk}}{10^{14}~{\rm cm^{-2}}}
\right)^{-0.75}~{\rm s}^{-1}\, ,
\label{eq:H2_diss}
\end{eqnarray}
where $I_{\rm IUV}$ is in cgs units.

For the \H2 formation on grains, recent experimental results have
indicated that $S\sim 0$ (the sticking efficiency of hydrogen
atoms;\ Appendix \ref{app:surface_rate}) for $T_{\rm dust}>20$ K
(Katz et al.\ 1999). Such a low-temperature threshold for the
\H2 formation suggests the following scenario. Because of the
thermal coupling between the dust and the cosmic microwave
background (CMB) of temperature $T_{\rm CMB}=2.7(1+z)$ K, molecular
formation on grains might have been strongly suppressed when
$T_{\rm CMB}\ga 20$ K or $z\ga 7$. As a result, galaxies start to
form stars actively when $z\sim 7$. In this case, $z\sim 7$ is the
typical redshift
for the onset of active star formation. Since such a typical
redshift, if it exists, could be in principle detected as an
enhancement of galaxy number counts at a certain flux level, it
is worth considering such a typical
``formation epoch'' in this paper. We reconsider this point in
detail in \S~\ref{sec:nc}.

SNe II affect the abundance of \H2 as they create regions
filled with hot gas ($T\sim 10^6$ K). In such an environment \H2
is destroyed, but it reforms after the gas cools; this
predominantly occurs in cooled shells (Shapiro \& Kang 1987;
Ferrara 1998; we note that our
$\fH2$ is roughly two times that of Shapiro \& Kang) and $\fH2$
becomes $\mbox{a few}\times 10^{-3}$. We cannot consider
these effects in this paper, because our model, which only
treats averaged quantities over the whole galaxy,
cannot describe local bubbles and shells produced by SNe II.
We expect that the formation of \H2 on dust surfaces and in
the cold shells keeps $\fH2$ from decreasing significantly.

In order to calculate the temperature evolution, cooling and
heating should be included in our model. We adopt the cooling
functions summarized in \S~2.3 of Hutchings et al.\ (2002),
i.e., cooling by molecular
hydrogen, and collisional excitation and ionisation of
atomic hydrogen. Again cooling by helium is neglected because
Hutchings et al.\ (2002) have shown that the temperature
evolution is little affected by the helium cooling. For the
heating by stellar IUV radiation,
we adopt equation (A21) of Kitayama \& Ikeuchi (2000).
The initial value for $T_{\rm gas}$ is assumed to be
$T_{\rm vir}$.

\subsection{Star formation law}\label{subsec:sfr}

The adopted star formation law is central to this paper,
because we propose a new ``paradigm'' based on the fact that
the stars are formed during the final cooling by molecular
hydrogen. If the molecular gas is abundant, stars are formed as
a result of
a dynamical collapse of gas. Since a representative timescale of
the dynamics of the gas disk
is the circular timescale, $t_{\rm cir}$, we expect that the star
formation rate $\psi$ is roughly
$M_{\rm gas}/t_{\rm cir}$ in a fully molecular gas.
Most of the ``semi-analytic'' recipes of galaxy evolution
(e.g., Kauffmann \& Charlot 1998; Somerville \& Primack 1999;
Cole et al.\ 2000; Granato et al.\ 2000; Nagashima et al.\ 2001;
cf.\ White \& Frenk 1991) assume this kind of law (the
dependence on the circular velocity is nonlinear though). However,
since the final coolant is molecular
hydrogen, we should include the hydrogen content in the formulation
to obtain a more physical star formation law. Therefore, we assume
the following form for the star formation rate:
\begin{eqnarray}
\psi (t)=f_{\rm H_2}(t)M_{\rm gas}/t_{\rm cir}\, ,\label{eq:sf_law}
\end{eqnarray}
where $M_{\rm gas}$ and $t_{\rm cir}$ are taken to be constant in
time.

Although the following results are critically dependent
on the assumed form of $\fH2$, the
experimental evidence shows that star formation depends positively
on molecular abundance (e.g., Rana \& Wilkinson 1986;
Wilson et al.\ 2000; Walter et al.\ 2002; but see e.g.,
Buat, Deharveng, \& Donas 1989; Tosi \& Diaz 1990).
Equation (\ref{eq:sf_law}) is given ``as a first
approximation'' in this paper to include this experimentally
supported law. Although multi-phase behavior of ISM can be also
important (McKee \& Ostriker 1977; Ikeuchi 1988;
Norman \& Spaans 1996),
this equation is expected to approximate the star formation rate
of the whole galaxy
in the framework of our one-zone treatment.

We are interested in objects whose virial temperature is
typically larger than $10^4$ K (\S~\ref{subsec:phys_state}).
Madau et al.\ (2001) have argued that such objects cool
on a timescale much shorter than the dynamical timescale and
consequently experience an initial strong episode of star
formation. Our approach in this paper
is conservative in the sense that active star formation does
not occur until a significant amount of hydrogen molecules is
produced. However, even in our ``conservative'' treatment, a
burst of
star formation occurs as a result of molecular formation on
dust grains (\S~\ref{sec:result}). Therefore, a scenario
similar to, albeit physically different from, those of
Madau et al.\ (2001), Ciardi et al.\ (2000), and Barkana (2002)
emerges from the present study.

\subsection{Evolution of dust content}\label{subsec:dust_formalism}

Because of short cosmic timescale between $z=5$ and 20 ($\la 1$ Gyr),
the contribution of Type Ia SN and winds from late-type stars to
dust formation is assumed to be negligible. In this case, SNe II are
the dominant sources for dust formation. The rate of SNe II as a
function of time, $\gamma (t)$, is given by
\begin{eqnarray}
\gamma (t)=\int_{8~M_\odot}^{\infty}
\psi (t-\tau_m)\, \phi (m)\, dm\, ,
\end{eqnarray}
where $\psi (t)$ is the star formation rate (SFR) at $t$ (for $t<0$,
$\psi (t)=0$),
$\phi (m)$ is the initial mass function (IMF; the definition of the
IMF is the same as that in Tinsley 1980), $\tau_m$ is the
lifetime of a star whose mass is $m$, and we assumed that
stars with $m>8~M_\odot$ produce SNe II. In this paper, we assume a
Salpeter IMF ($\phi (m)\propto m^{-2.35}$) with the stellar mass
range of 0.1--60 $M_\odot$. It has been suggested that the
IMF is much more weighted to massive stars, i.e., top-heavy,
in primeval galaxies (e.g., Bromm et al.\ 2002).
In a top-heavy environment, the production of dust by massive
stars is enhanced. As a result,
dust amount, and thus molecular amount, in galaxies would be
larger than that predicted in this paper. Here, we
``conservatively'' assume the Salpeter IMF.

Dust destruction by SNe II can be important. The destruction
timescale $\tau_{\rm SN}$ is estimated to be (McKee 1989;
Lisenfeld \& Ferrara 1998)
\begin{eqnarray}
\tau_{\rm SN} =
\frac{M_{\rm g}}{\gamma\epsilon M_{\rm s}(100~{\rm km~s}^{-1})}
\, ,
\end{eqnarray}
where $M_{\rm s}(100~{\rm km~s}^{-1})=6.8\times 10^3~M_\odot$
(Lisenfeld \& Ferrara 1998) is the mass accelerated to
100 km s$^{-1}$ by a SN blast,
$\gamma$ is the SN II rate, $\epsilon\sim 0.1$ (McKee 1989) is the
efficiency of dust destruction in a medium shocked by a SN II.
Since we are interested in the first star formation activity,
we assume the relation between stellar mass and lifetime of
zero-metallicity stars in Table 6 of Schaerer (2002) (the case
without mass loss is applied).

Then the rate of increase of $M_{\rm d}$ is written as
\begin{eqnarray}
\frac{d{M}_{\rm d}}{dt}=m_{\rm d}\gamma-
\frac{{M}_{\rm d}}{\tau_{\rm SN}}\, ,
\end{eqnarray}
where $m_{\rm d}$ is the typical dust mass produced in a SN II.
Todini \& Ferrara (2001) showed that $m_{\rm d}$ varies
with progenitor mass and metallicity. There is also some
uncertainty in the explosion energy of a SN II.
The Salpeter IMF-weighted mean of dust mass produced per SN II
for the 1) $Z=0$, Case A, 2) $Z=0$, Case B, 3) $Z=10^{-2}Z_\odot$,
Case A, and 4) $Z=10^{-2}Z_\odot$,
Case B are 1) 0.22 $M_\odot$, 2) 0.46  $M_\odot$, 3) 0.45
$M_\odot$, and 4) 0.63  $M_\odot$, respectively
($Z$ is the metallicity, and Cases A and B correspond to
low and high explosion energy\footnote{The kinetic energies given to
the ejecta are $\sim 1.2\times 10^{51}$ ergs and $\sim 2\times 10^{51}$
ergs for Case A and Case B, respectively.}, respectively). We adopt the
average of the four cases, i.e., $m_{\rm d}\simeq 0.4~M_\odot$ but
we should remember that $m_{\rm d}$ can have a range (from 0.22
to 0.63 $M_\odot$). Since the dust destruction is negligible as shown
later, our final value of the dust mass is approximately
proportional to the adopted one for $m_{\rm d}$.

\subsection{Evolution of metal content}

The evolution of metal content can be predicted once the
star formation history and metal yield per SN II are fixed
(e.g., Tinsley 1980). Suginohara, Suginohara \& Spergel (1999)
have proposed that the fluctuations in space and wavelength
of (sub-)millimetre
background radiation made of high-redshift metal lines can be
used as an indicator of structure formation at high redshift.
Here we obtain the metal mass injected in the gas phase by
subtracting the dust mass from the metal mass.
The evolution of the mass of a heavy element (species $i$) in
the gas phase of a galaxy, $M_i$, is thus calculated by
\begin{eqnarray}
\frac{dM_i}{dt}=m_i\gamma -\frac{dM_{{\rm dust},i}}{dt}\, ,
\label{eq:metal}
\end{eqnarray}
where $m_i$ is the averaged mass of element $i$ formed per
SN II and $M_{{\rm dust},i}$ is the mass of element $i$ in the
dust phase. According to Todini \& Ferrara's calculation, dust
contains 15\% of oxygen and 36\% of carbon. Therefore, we
assume that
$M_{\rm dust,O}=0.15M_{\rm dust}$ and that
$M_{\rm dust,C}=0.36M_{\rm dust}$ (we take the same average
as that in \S~\ref{subsec:dust_formalism}). In
Table \ref{tab:metal}, we list $m_i$ for carbon and oxygen.
When calculating $m_i$, we adopted the results by
Woosley \& Weaver (1995) and took the same mean as that in
\S~\ref{subsec:dust_formalism}. Suginohara et al.\ (1999) also
considered nitrogen, but we have not included this species as
the production of nitrogen in a SN II is one or two orders of
magnitude less than that of oxygen or carbon. 

\subsection{Radiative properties}\label{subsec:luminosities}

\subsubsection{UV and FIR}\label{subsubsec:UV_FIR}

Probably the most direct way to reveal high-redshift dust is
to observe FIR emission. We now derive the evolution of
galactic FIR luminosity and dust temperature. Because of the
large cross section of dust against IUV light and the intense
IUV radiation field in a star-forming galaxy, we can assume that
the FIR luminosity is equal to the absorbed energy of
IUV light.

First, we should estimate the fraction of the IUV radiation absorbed
by dust. For the convenience of the following calculation, we
define the following typical optical depth in the vertical
direction of the disk, $\tau_0$, as
\begin{eqnarray}
\tau_{0}\equiv\pi a^2Q_{\rm UV}n_{\rm dust}H\, .\label{eq:def_tau0}
\end{eqnarray}
In this paper we assume single value for $a$, because there is a
typical size of dust produced by SNe II as shown by
Todini \& Ferrara (2001). We note that this optical
depth is different from $\tau_{\rm dust}$ in equation
(\ref{eq:UVnormalize}), where we needed an optical depth in the
disk direction. Here, the optical depth in the vertical
direction is useful as we see in the following.

Since $\tau_0$ is independent of $H$, the optical depth of the
dust does not depend on the treatment of
$r_{\rm disk}/H$ in \S~\ref{subsubsec:density}. The dust density is
related to the mean dust number density as
\begin{eqnarray}
\frac{4\pi}{3}a^3\delta n_{\rm dust}\pi r_{\rm disk}^22H=
M_{\rm dust}\, ,\label{eq:dust_density}
\end{eqnarray}
where $\delta$ is the grain material density. By solving
equation (\ref{eq:dust_density}) for $n_{\rm dust}$ and
substituting it into equation (\ref{eq:def_tau0}), we obtain
\begin{eqnarray}
\tau_0=\frac{3}{8\pi}
\frac{Q_{\rm UV}M_{\rm d}}{a\delta r_{\rm disk}^2}\, ,
\end{eqnarray}
If the angle between the direction of a IUV light propagation and
the vertical direction of the disk is $\theta$, the intensity of
the light becomes roughly $\exp (-\tau_0/\cos\theta )$.
Therefore, the luminosity of IUV light escaping from the galactic
disk, $L_{\rm UV}$ is estimated to be
\begin{eqnarray}
L_{\rm UV}\simeq L_{\rm UV,0}\left\langle\exp\left(-
\frac{\tau_0}{\cos\theta}\right)\right\rangle_\theta
=L_{\rm UV,0}\, E_2(\tau_0)\, ,
\end{eqnarray}
where $\langle\cdot\rangle_\theta$ indicates the mean over the
solid angle and $ L_{\rm UV,0}$ is the intrinsic UV luminosity
of the galaxy. The exponential integral $E_n(x)$ ($n=0,$ 1, 2, ...;
$x>0$) is defined as
\begin{eqnarray}
E_n(x)\equiv x^{n-1}\int_x^\infty\frac{\exp (-y)}{y}\, dy\, .
\end{eqnarray}
$L_{\rm UV,0}$ is assumed to be equal to the total luminosity of
OB stars whose mass is larger than 3 $M_\odot$ (Cox 2000):
\begin{eqnarray}
L_{\rm UV,0}(t)=\int_{3~M_\odot}^{\infty}dm\,
\int_0^{\tau_m}dt'\, L(m)\,\phi (m)\,\psi (t-t')\, ,\label{eq:uv_lum}
\end{eqnarray}
where $L(m)$ is the stellar luminosity as a function of stellar
mass ($m$). For $L(m)$, we adopt the model of zero-metallicity stars
without mass loss in Schaerer (2002). We fix this relation
as a first approximation in this paper.

We adopt $a\simeq 0.03~\mu$m (Todini \& Ferrara 2001),
$Q_{\rm UV}\simeq 1$, and $\delta\simeq 2$ g cm$^{-3}$
(Draine \& Lee 1984). We only consider a single value of $a$,
because there is a well defined sharp peak in the dust size
distribution by Todini \& Ferrara (2001). Then we obtain the
energy absorbed by dust. We assume that all the absorbed energy
is reemitted in the FIR. Thus, the FIR luminosity $L_{\rm FIR}$
becomes
\begin{eqnarray}
L_{\rm FIR}=L_{\rm UV,0}-L_{\rm UV}\, .
\end{eqnarray}

The dust temperature $T_{\rm dust}$ is determined from the equation
derived by Hirashita et al.\ (2002a) based on the dust emissivity
given by Draine \& Lee (1984) for $a=0.03~\mu$m:
\begin{eqnarray}
T_{\rm d}=20\left(
\frac{L_{\rm FIR}/L_\odot}{2.5\times 10^2M_{\rm d}/M_\odot}
\right)^{1/6}~{\rm K}\, .
\end{eqnarray}

\subsubsection{Metal lines}\label{subsubsec:metal}

We next calculate the luminosities of the metal lines. Our
model calculation indicates that almost all the gas is in
neutral form ($x\ll 1$ and $\fH2\ll 1$); hence we consider lines
typical for neutral regions. The most interesting lines are
listed in
Table \ref{tab:metal}. For spontaneous transition from an
upper level $u$ to a lower level $l$, the total luminosity of
the line with
frequency $\nu_{\rm line}$ ($\mbox{line}=\mbox{C609}$,
C370, O63, O146 for C {\sc i} 609 $\mu$m, C {\sc i} 370 $\mu$m,
O {\sc i} 63.2 $\mu$m, and O {\sc i} 146 $\mu$m, respectively)
becomes
\begin{eqnarray}
L_{\rm line}=h_{\rm P}\nu_{\rm line}{\cal N}_uA_{ul}\, ,
\end{eqnarray}
where $h_{\rm P}$ is the Planck constant, ${\cal N}_u$
is the total number of the atoms in the upper level in the
whole galaxy, and $A_{ul}$ is the Einstein coefficient of
the spontaneous emission. We assume that all the carbon and
the oxygen are neutral; we will also consider the
C {\sc ii} 158 $\mu$m line, which is also emitted from the
neutral medium (Tielens \& Hollenbach 1985; Liszt 2002). If the
timescale of
the spontaneous emission is longer than that of the collisional
excitation, the fraction within the state is proportional to
$2g+1$ (i.e., 1/9, 3/9, and 5/9 for
$^3P_0$, $^3P_1$, and $^3P_2$, respectively).
However, if the density is so low that the
collisional excitation does not occur so frequently as
the spontaneous emission, the population in the upper level
is reduced. We include this effect of low density by
multiplying the factor ${\cal F}$ described in
Appendix \ref{app:line}.

The line intensity is estimated for the case where
the population ratio is determined by $2g+1$. Since this gives
us the maximum line intensity, we denote the intensity as
$L_{\rm line}^{\rm max}$; more realistically, the line
intensity is closer to ${\cal F}L_{\rm line}^{\rm max}$. We
estimate ${\cal N}_u$ from the metal mass in
equation (\ref{eq:metal}) and the number fraction in the
upper level. The actual intensity of metal line is reduced by
a factor ${\cal F}$. However, a precise determination of
${\cal F}$ requires a detailed treatment of the evolution of
gas density. As discussed in \S~\ref{subsubsec:density}, our
present framework can only provide a first order estimate of
such quantity; therefore, only $L_{\rm line}^{\rm max}$ is
calculated in this paper. Nevertheless we propose a simple
method to estimate ${\cal F}$ in Appendix \ref{app:line}.
When we calculate the C {\sc ii} 158 $\mu$m line intensity,
we assume that all the carbon atoms are in  the form of C$^+$.

\section{RESULTS FOR FIDUCIAL GALAXIES}\label{sec:result}

\subsection{A typical primeval galaxy}

In the following we show the evolution of some characteristic
quantities predicted by our model for a galaxy with
$(\Mvir ,\,\zvir )=(10^9~M_\odot ,\, 10)$, corresponding to a
2.5-$\sigma$ density fluctuation of the cosmological density
field. The qualitative behavior of the quantities are similar
for other objects of interest to the present study. In the
next subsection, we explore various $(\Mvir ,\,\zvir )$ values.

\begin{figure}
\includegraphics[width=8.5cm]{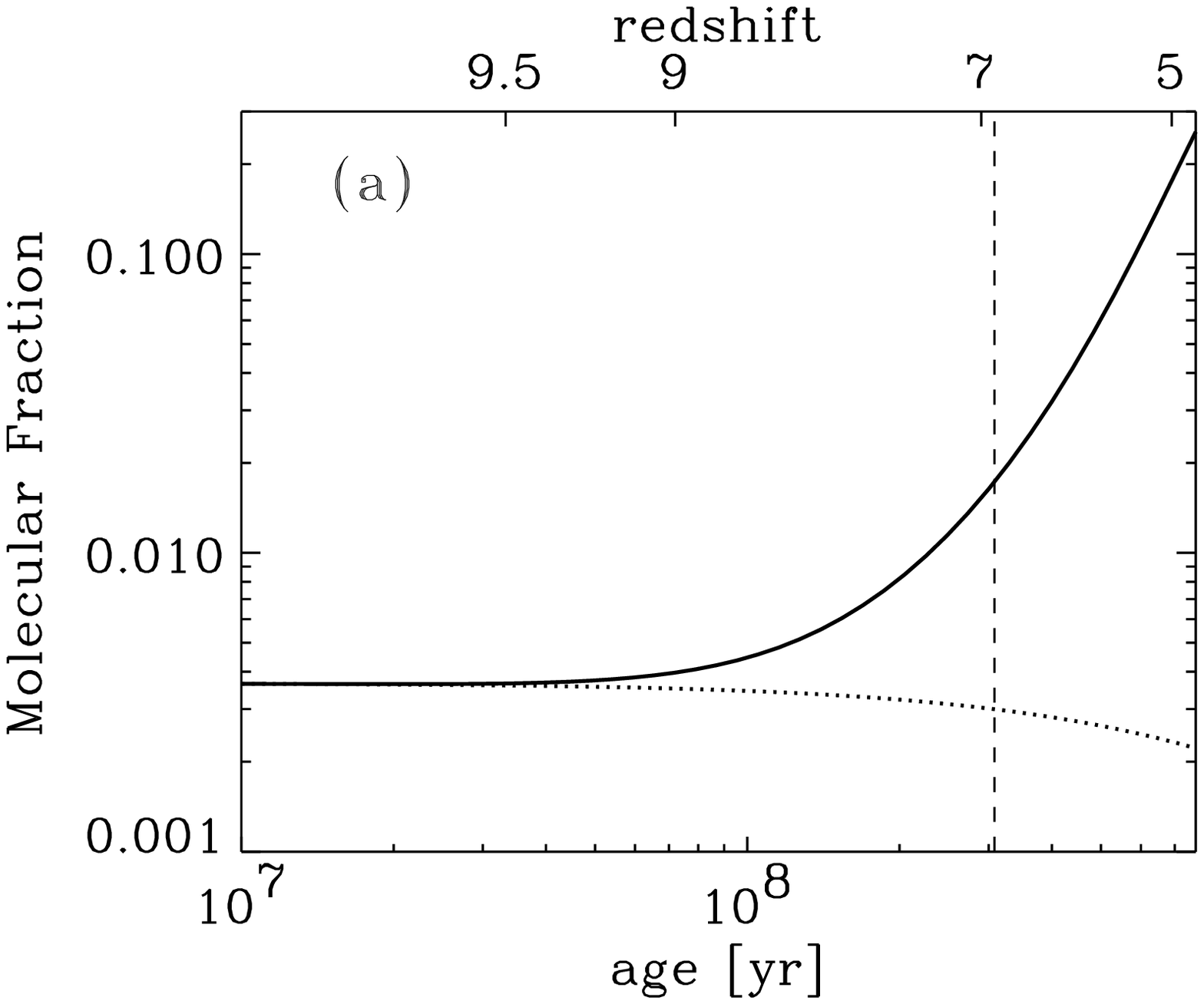}
\includegraphics[width=8.5cm]{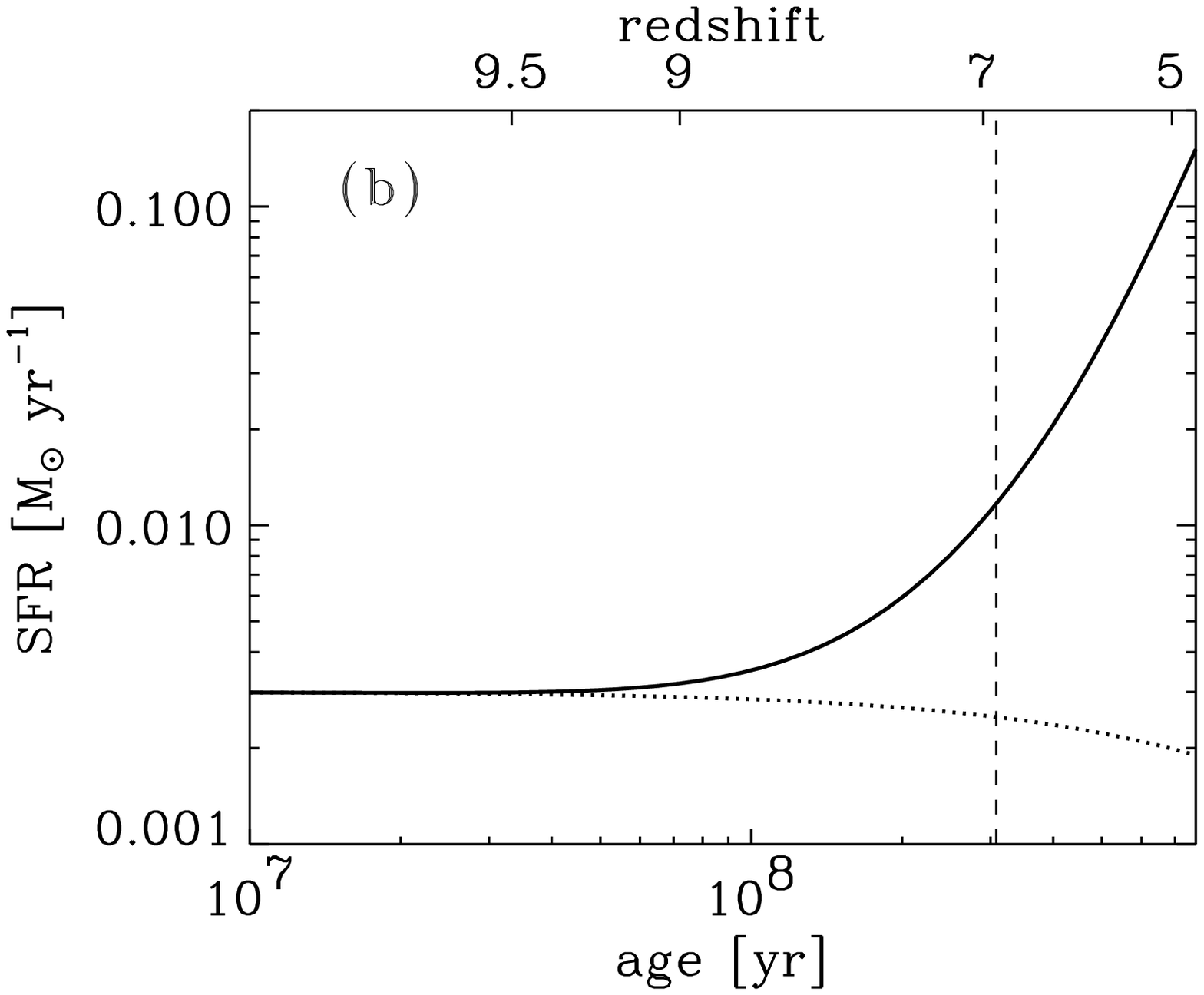}
\caption{Time evolution of (a) the molecular fraction and
(b) the star formation rate for a galaxy with $\zvir=10$ and
$\Mvir =10^9~M_\odot$ corresponding to the 2.5-$\sigma$
density fluctuation. The evolution in the first 10 circular
times is shown. The solid line is for the case where we consider
dust (and therefore the H$_2$ formation on grains), and the dotted
line is for the case of no dust production ($M_{\rm dust}=0$).
The vertical dashed line marks the 4 circular timescales,
where the typical quantities are defined in the text and in
Table \ref{tab:sfr}. In the top axis, we show the corresponding
redshifts.
\label{fig:H2}}
\end{figure}

\begin{table*}
\begin{center}
\caption{Typical quantities for various sets of $(\Mvir, \zvir )$}
$\zvir =20$ \\
\begin{tabular}{lccccccccc}\hline
$\Mvir$ & $\bar{\psi}$ & $\bar{L}_{\rm FIR}$ & $\bar{L}_{\rm UV}$ &
$\bar{L}_{\rm C609}^{\rm max}$ & $\bar{L}_{\rm C370}^{\rm max}$ &
$\bar{L}_{\rm O63}^{\rm max}$ &
$\bar{L}_{\rm O146}^{\rm max}$ & $\bar{L}_{\rm C158}^{\rm max}$ &
$\bar{T}_{\rm dust}$ \\
($M_\odot$) & ($M_\odot~{\rm yr}^{-1}$) & ($L_\odot$) & ($L_\odot$) &
($L_\odot$) & ($L_\odot$) & ($L_\odot$) & ($L_\odot$)  & ($L_\odot$) &
(K) \\
\hline
$10^8$ & 2.5e$-$3 & 2.5e6 & 4.0e6 & 4.9e2 & 5.3e1 & 1.7e7 & 4.2e5
& 6.1e3 & 44 \\
$10^9$ & 1.9e$-$1 & 4.0e8 & 4.6e7 & 2.1e4 & 2.2e3 & 6.1e8 & 1.5e7
& 2.6e5 & 57 \\
$10^{10}$ & 4.4e0 & 1.0e10 & 1.4e7 & 4.5e5 & 4.9e4 & 1.1e10 & 2.7e8
& 5.6e6 & 61 \\
$10^{11}$ & 4.3e1 & 9.8e10 & 4.2e5 & 4.3e6 & 4.7e5 & 1.1e11 & 2.7e9
& 5.4e7 & 61 \\
\hline
\end{tabular}
\\ ~ \\
$\zvir =15$ \\
\begin{tabular}{lccccccccc}\hline
$\Mvir$ & $\bar{\psi}$ & $\bar{L}_{\rm FIR}$ & $\bar{L}_{\rm UV}$ &
$\bar{L}_{\rm C609}^{\rm max}$ & $\bar{L}_{\rm C370}^{\rm max}$ &
$\bar{L}_{\rm O63}^{\rm max}$ &
$\bar{L}_{\rm O146}^{\rm max}$ & $\bar{L}_{\rm C158}^{\rm max}$ &
$\bar{T}_{\rm dust}$ \\
($M_\odot$) & ($M_\odot~{\rm yr}^{-1}$) & ($L_\odot$) & ($L_\odot$) &
($L_\odot$) & ($L_\odot$) & ($L_\odot$) & ($L_\odot$)  & ($L_\odot$) &
(K) \\
\hline
$10^8$ & 9.5e$-$4 & 6.7e5 & 2.1e6 & 4.0e2 & 4.3e1 & 1.4e7 & 3.5e5
& 5.0e3 & 37 \\
$10^9$ & 7.3e$-$2 & 1.4e8 & 5.1e7 & 1.8e4 & 1.9e3 & 5.2e8 & 1.3e7
& 2.2e5 & 49 \\
$10^{10}$ & 2.7e0 & 6.8e9 & 5.2e7 & 5.9e5 & 6.4e4 & 1.4e10 & 3.3e8
& 7.3e6 & 55 \\
$10^{11}$ & 2.4e1 & 6.0e10 & 1.5e7 & 4.9e6 & 5.3e5 & 1.2e11 & 2.9e9
& 6.0e7 & 55 \\
\hline
\end{tabular}
\\ ~ \\
$\zvir =12$ \\
\begin{tabular}{lccccccccc}\hline
$\Mvir$ & $\bar{\psi}$ & $\bar{L}_{\rm FIR}$ & $\bar{L}_{\rm UV}$ &
$\bar{L}_{\rm C609}^{\rm max}$ & $\bar{L}_{\rm C370}^{\rm max}$ &
$\bar{L}_{\rm O63}^{\rm max}$ &
$\bar{L}_{\rm O146}^{\rm max}$ & $\bar{L}_{\rm C158}^{\rm max}$ &
$\bar{T}_{\rm dust}$ \\
($M_\odot$) & ($M_\odot~{\rm yr}^{-1}$) & ($L_\odot$) & ($L_\odot$) &
($L_\odot$) & ($L_\odot$) & ($L_\odot$) & ($L_\odot$)  & ($L_\odot$) &
(K) \\
\hline
$10^8$ & 4.0e$-$4 & 1.7e5 & 1.0e6 & 2.8e2 & 3.0e1 & 1.0e7 & 2.5e5
& 3.5e3 & 31 \\
$10^9$ & 2.7e$-$2 & 3.8e7 & 3.7e7 & 1.0e4 & 1.1e3 & 3.4e8 & 8.2e6
& 1.3e5 & 42 \\
$10^{10}$ & 1.3e0 & 3.3e9 & 2.0e8 & 4.2e5 & 4.5e4 & 1.1e10 & 2.6e8
& 5.2e6 & 51 \\
$10^{11}$ & 1.2e1 & 3.0e10 & 2.5e8 & 3.5e6 & 3.7e5 & 9.1e10 & 2.2e9
& 4.3e7 & 51 \\
\hline
\end{tabular}
\\ ~ \\
$\zvir =9$ \\
\begin{tabular}{lccccccccc}\hline
$\Mvir$ & $\bar{\psi}$ & $\bar{L}_{\rm FIR}$ & $\bar{L}_{\rm UV}$ &
$\bar{L}_{\rm C609}^{\rm max}$ & $\bar{L}_{\rm C370}^{\rm max}$ &
$\bar{L}_{\rm O63}^{\rm max}$ &
$\bar{L}_{\rm O146}^{\rm max}$ & $\bar{L}_{\rm C158}^{\rm max}$ &
$\bar{T}_{\rm dust}$ \\
($M_\odot$) & ($M_\odot~{\rm yr}^{-1}$) & ($L_\odot$) & ($L_\odot$) &
($L_\odot$) & ($L_\odot$) & ($L_\odot$) & ($L_\odot$)  & ($L_\odot$) &
(K) \\
\hline
$10^8$ & 1.4e$-$4 & 3.1e4 & 4.2e5 & 1.8e2 & 1.9e1 & 6.5e6 & 1.6e5
& 2.2e3 & 25 \\
$10^9$ & 7.9e$-$3 & 6.5e6 & 1.7e7 & 5.9e3 & 6.3e2 & 2.0e8 & 4.9e6
& 7.2e4 & 34 \\
$10^{10}$ & 5.3e$-$1 & 1.2e9 & 3.3e8 & 2.8e5 & 3.0e4 & 7.6e9 & 1.8e8
& 3.4e6 & 45 \\
$10^{11}$ & 4.6e0 & 1.2e10 & 1.1e9 & 2.3e6 & 2.5e5 & 6.5e10 & 1.6e9
& 2.8e7 & 46 \\
\hline
\end{tabular}
\\ ~ \\
$\zvir =6$ \\
\begin{tabular}{lccccccccc}\hline
$\Mvir$ & $\bar{\psi}$ & $\bar{L}_{\rm FIR}$ & $\bar{L}_{\rm UV}$ &
$\bar{L}_{\rm C609}^{\rm max}$ & $\bar{L}_{\rm C370}^{\rm max}$ &
$\bar{L}_{\rm O63}^{\rm max}$ &
$\bar{L}_{\rm O146}^{\rm max}$ & $\bar{L}_{\rm C158}^{\rm max}$ &
$\bar{T}_{\rm dust}$ \\
($M_\odot$) & ($M_\odot~{\rm yr}^{-1}$) & ($L_\odot$) & ($L_\odot$) &
($L_\odot$) & ($L_\odot$) & ($L_\odot$) & ($L_\odot$)  & ($L_\odot$) &
(K) \\
\hline
$10^8$ & 4.1e$-$5 & 3.4e3 & 1.3e5 & 1.0e2 & 1.1e1 & 3.9e6 & 9.5e4
& 1.3e3 & 19 \\
$10^9$ & 1.8e$-$3 & 5.9e5 & 4.9e6 & 3.0e3 & 3.3e2 & 1.1e8 & 2.7e6
& 3.8e4 & 25 \\
$10^{10}$ & 1.3e$-$1 & 1.7e8 & 2.2e8 & 1.4e5 & 1.5e4 & 4.2e9 & 1.0e8
& 1.7e6 & 36 \\
$10^{11}$ & 1.4e0 & 2.8e9 & 1.5e9 & 1.4e6 & 1.5e5 & 4.3e10 & 1.1e9
& 1.7e7 & 39 \\
\hline
\end{tabular}
\label{tab:sfr}
\end{center}
\end{table*}

Using the equations above, we simultaneously calculate the time
evolution of $f_{\rm H_2}$, $\psi$, $M_{\rm dust}$, and various
kinds of luminosities in a self-consistent manner. In
Fig.\ \ref{fig:H2}, we show the evolution of the molecular fraction
$f_{\rm H_2}$ (solid line of Fig.\ \ref{fig:H2}a), and the star
formation rate $\psi$ (solid line of Fig.\ \ref{fig:H2}b) for an
object $\zvir=10$
and $\Mvir =10^9~M_\odot$. We show the results up to
$10t_{\rm cir}$ ($\simeq 7.8\times 10^8$ yr), which is
comparable to the time
between $z=10$ and $z=5$. At $t=10t_{\rm cir}$, 
$\sim 30$\% of the gas is converted into stars. Therefore, the
calculation further than this time is not consistent with our
assumptions, which neglect the gas conversion into stars. The
gas temperature
rapidly drops and reaches $\la 300$ K within $10^7$ yr.

For other sets of $(\Mvir ,\,\zvir )$, an enhancement of the
star formation at 3--5$t_{\rm cir}$ is commonly seen. This results
from the H$_2$ formation on dust grains. The dotted lines of
Figs.\ \ref{fig:H2}a and \ref{fig:H2}b indicate the case in which
we neglect the dust (i.e., $M_{\rm d}=0$) at all times. In this
case, after stars form, $\fH2$ decreases because of
photo-dissociation. Comparing the solid and dotted lines in
Figs.\ \ref{fig:H2}a and b, we conclude that the existence of
dust is essential in causing a strong star formation activity in
primeval galaxies.

\begin{figure}
\begin{center}
\includegraphics[width=8.5cm]{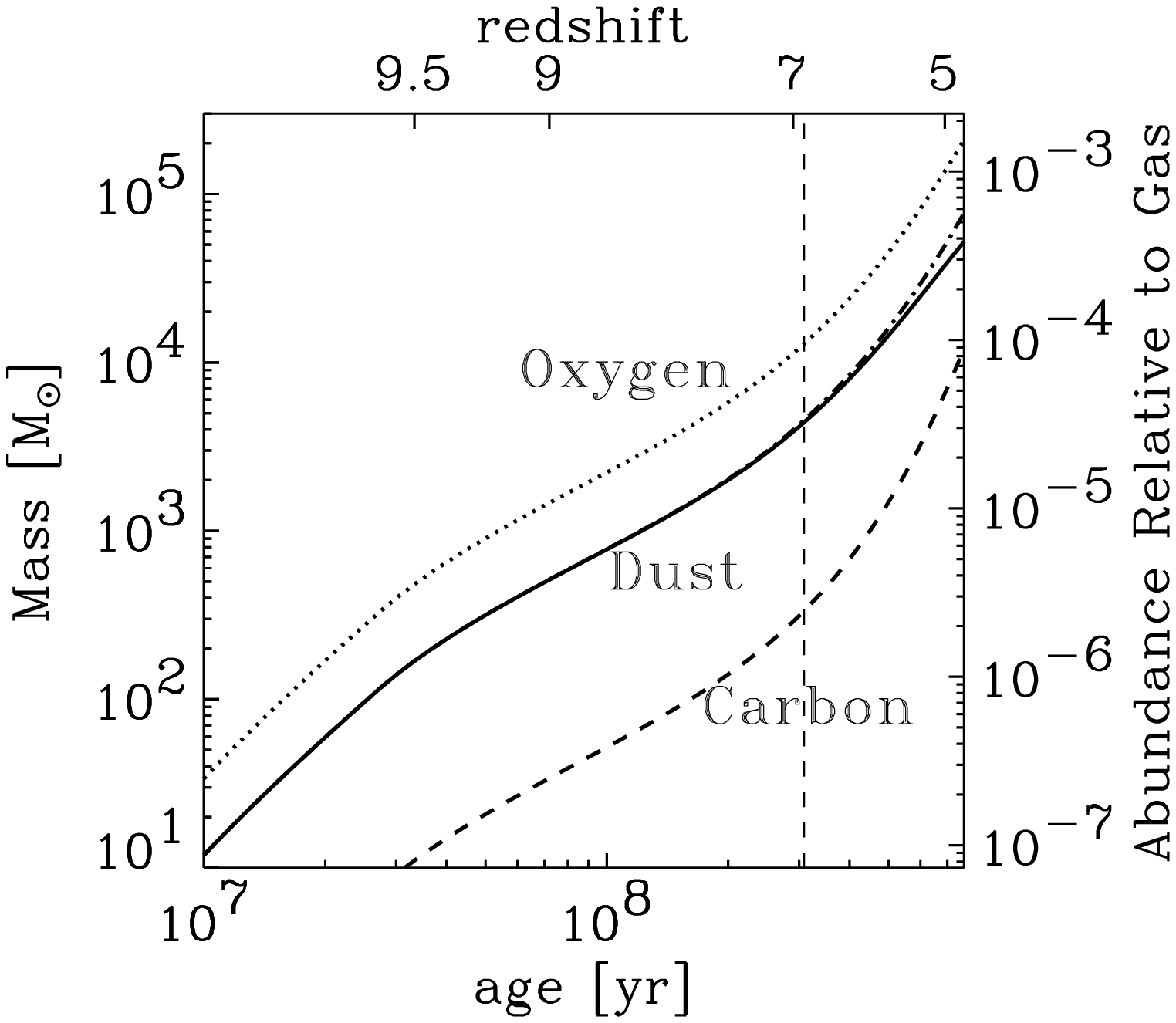}
\end{center}
\caption{Time evolution of the masses of dust, oxygen (in gas
phase), and carbon (in gas phase) in the same galaxy as
Fig.\ \ref{fig:H2} (solid, dotted, and dashed lines,
respectively). The dot-dashed
line represents the case in which dust destruction is neglected.
The vertical dashed line shows the
4 circular timescales.\label{fig:dust_ev}}
\end{figure}

In Figure \ref{fig:dust_ev}, we show
the evolution of dust, oxygen and carbon (solid, dotted, and
dashed lines, respectively). The mass fraction relative to
the total gas mass is also shown on the right axis.
Note that the solar abundances for oxygen and carbon
are $1.0\times 10^{-2}$ and $3.4\times 10^{-3}$ in mass ratio,
respectively (Anders \& Grevesse 1989; Cox 2000). We see
rapid rises of those masses around
$t\sim 4t_{\rm cir}$ (marked with the vertical dashed
line). We also show the case in which dust destruction
by SNe II is neglected (dot-dashed line): dust destruction has
little influence on the dust amount during the early
($\la\mbox{several}~t_{\rm cir}$) evolution.

\begin{figure}
\includegraphics[width=8.5cm]{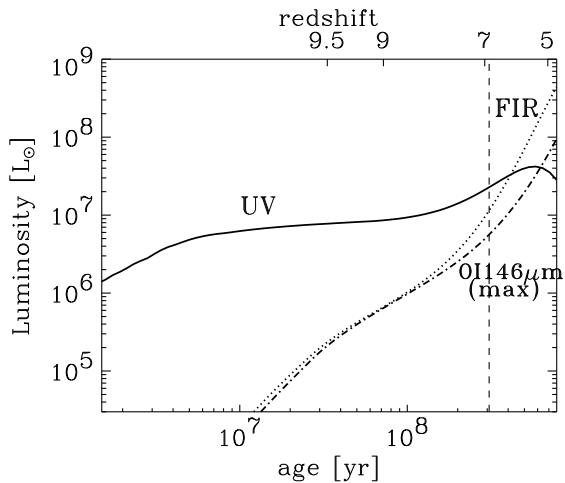}
\caption{Time evolution of the ultraviolet luminosity
(solid line), the far-infrared luminosity (dotted line), and
the maximum line luminosity of
O {\sc i} 146 $\mu$m ($L_{\rm O146}^{\rm max}$)
(dot-dashed line) for the same galaxy as in
Fig.\ \ref{fig:H2}. The vertical dashed line shows the
4 circular timescales.\label{fig:fir_luminosity}}
\end{figure}

In Figure \ref{fig:fir_luminosity}, we show the evolution of
three luminosities; $L_{\rm UV}$, $L_{\rm FIR}$, and
$L_{\rm O146}^{\rm max}$ (solid, dotted, and dot-dashed
lines, respectively). Whereas $L_{\rm FIR}$ and
$L_{\rm O146}^{\rm max}$ grow rapidly around
$t\sim 4t_{\rm cir}$, $L_{\rm UV}$
does not increase as significantly as the other two
luminosities because of dust extinction. The FIR luminosity
becomes comparable to the UV luminosity soon after the active
star formation phase starts. Therefore, even in the early
stages of star formation, dust absorbs a significant
fraction of stellar light and reprocesses it to the FIR range.
This is because of the dense and compact nature of
high-redshift galaxies ($r_{\rm vir}\propto (1+z)^{-1}$ with a
fixed $\Mvir$). Therefore, even at early cosmic epochs ($z>5$)
simultaneous observations of FIR emission from dust and of the
UV/NIR stellar light is crucial to trace the total radiative
energy from galaxies.

\subsection{Dependence on $(\Mvir,\,\zvir)$}

Table \ref{tab:sfr} clarifies the dependence of various quantities
on $\Mvir$ and $\zvir$. We define the typical star formation rate
for each object, $\bar{\psi}(\Mvir ,\, \zvir)$, as
$\psi (4t_{\rm cir})$. This is because after 4 circular
times objects have consumed $\sim 5$\% of the gas. Since it is
empirically known that the star formation efficiency during an
episode of star formation activity is $\la 10$\%
(e.g., Inoue et al.\ 2000; Barkana 2002), the definition yields
a good working value. Moreover, the value at $t=4t_{\rm cir}$
gives a good
average for the star formation rates over the Hubble timescale
at $\zvir$ within a factor of $\sim 2$. In Table \ref{tab:sfr},
we show the typical
star formation rate as a function of $\zvir$ and $\Mvir$. In
Table \ref{tab:sfr} we
also show other quantities,
all of which are estimated at $t=4t_{\rm cir}$ (i.e., the
quantities with the upper bars indicates the typical quantities
estimated at $t=4t_{\rm cir}$).

All quantities except for $\bar{L}_{\rm UV}$ are monotonically
increasing functions of both $\Mvir$ and $\zvir$. An object with
larger $\Mvir$ tends to contain a larger gas mass ($M_{\rm gas}$).
An object with larger $\zvir$ tends to have a short $t_{\rm cir}$
($\propto (1+\zvir )^{-1.5}$)
because of a small $r_{\rm disk}\propto (1+\zvir )^{-1}$ and a
large $v_{\rm c}\propto (1+\zvir )^{0.5}$. Therefore, the star
formation law (eq.\ \ref{eq:sf_law})
indicates that $\bar{\psi}$ becomes an
increasing function of both $\Mvir$ and $\zvir$. The larger
the star formation rate is, the larger the
metal production, the dust production, and the intrinsic
OB star luminosity. As a result almost all quantities
become an increasing function of both $\Mvir$ and $\zvir$.
On the contrary, $\bar{L}_{\rm UV}$ is not monotonic. For
objects with $\zvir\ga 15$ and $\Mvir\ga 10^{10}~M_\odot$,
$\bar{L}_{\rm UV}$ decreases as $\Mvir$ and $\zvir$ increase.
This is because of the compactness of high-redshift objects.
Dust accumulation in a compact region makes the optical depth
of dust very large; hence UV radiation is efficiently absorbed
by dust. However, objects with $\zvir\ga 15$ and
$\Mvir\ga 10^{10}~M_\odot$ are extremely rare because they
correspond to density fluctuations $\ga 4~\sigma$.

The temperature of dust is generally higher than that
observed in normal spiral galaxies ($\sim 20$ K), because
the compactness (meaning high density) of the high-redshift
galaxies leads to a strong stellar radiation field. For a
nearby dwarf galaxy, SBS 0335$-$052, the compactness of the
star-forming region is also suggested to result in a high
dust temperature (Dale et al.\ 2001; Hirashita et al.\ 2002a).

\section{Statistical Predictions}\label{sec:nc}

Future observations will provide large samples of galaxies at
$z>5$. Therefore, predictions of the statistical
properties of galaxies in such a redshift range are very
valuable. Using the result above, we focus on cosmic star
formation rate (Tinsley \& Danly 1980; Madau et al.\ 1996),
number counts of galaxies,  integrated light
(Partridge \& Peebles 1967), where ``integrated'' means the
sum of the contribution from all the galaxies of interest.

\subsection{Cosmic Star Formation History}

The star formation rate per unit comoving volume at the redshift
$z$, $\Psi (z)$, is written as
\begin{eqnarray}
\Psi (z)=\int_{M_{\rm min}}^{M_{\rm max}}dM\, \psi (M,\, z)\,
\frac{\partial n(M,\, z)}{\partial M}\, ,\label{eq:cosmic_sfh}
\end{eqnarray}
where $n(M,\, z)$ is the comoving number
density of halos with masses larger than $M$ at $z$, and
$\psi (M,\, z)$ is the star formation
rate of an object with mass $M$ at $z$.

In equation (\ref{eq:cosmic_sfh}), $M_{\rm min}$ is set equal
to the maximum mass of a galaxy whose gas is blown away by
SNe II in order to exclude objects
which do not form stars continuously. From a fitting to the
result of Ciardi et al.\ (2000), we adopt
\begin{eqnarray}
M_{\rm min}=\frac{5\times 10^{10}}{(1+z)^{1.5}}~M_\odot\, .
\label{eq:benedetta}
\end{eqnarray}
Ciardi et al.\ presented three cases (A, B, and C) for various
cooled gas fraction, star formation efficiency, and escaped
photon fraction, all of which are inherent to their model. In
equation (\ref{eq:benedetta}), we adopt their
fiducial case A. Even for the other cases, the following results
are not changed at all, because
galaxies whose mass is as low as $10^9~M_\odot$ do not
contribute to the number counts at the flux level of interest.
Ciardi et al.'s case A indicates that the 1.2\%
($=f_{\rm b}f_\star$) of the baryons are converted into stars
in a dynamical time. This gives a star
formation rate comparable to the one that we find at
$t=4t_{\rm cir}$.

For the maximum mass we adopt $M_{\rm max}=10^{13}~M_\odot$.
As long as $M_{\rm max}\ga 10^{13}~M_\odot$, the results are
not changed at all, because the number of objects with
$\Mvir >10^{13}~M_\odot$ is
negligible in all the considered redshift range ($5<z<20$). To
derive $\partial n(M,\, z)/\partial M$, we use the
Press-Schechter formalism (Press \& Schechter 1974). We
assume that the star formation rate is a function of the virial
mass and the redshift as listed in Table \ref{tab:sfr};
i.e., $\psi (M,\, z)=\bar{\psi}(M_{\rm vir}=M\, , \zvir =z)$.

\begin{figure}
\includegraphics[width=8.5cm]{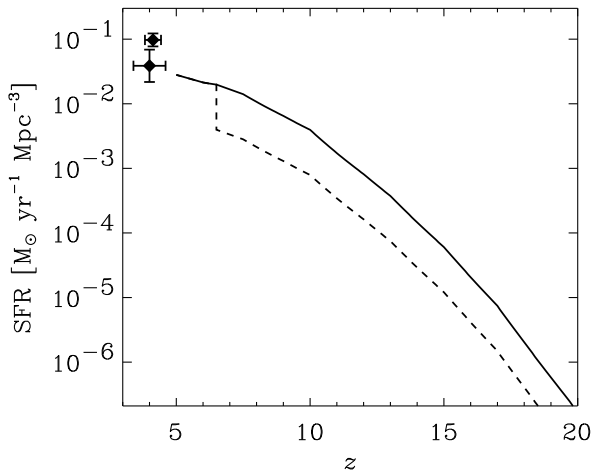}
\caption{Star formation rate per unit comoving volume (solid line).
The dotted line indicates the star formation history with a
``burst'' (bSFH) at $z=6.5$. The two data points around $z\sim 4$
are from Madau et al.\ (1998)
(lower) and Steidel et al.\ (1999) (upper). These points have been
corrected for the dust extinction by Steidel et al.\ (1999) and
for the cosmological parameters chosen in this paper.
\label{fig:cosm_sfh}}
\end{figure}

In Figure \ref{fig:cosm_sfh}, we show $\Psi (z)$ (solid line).
The considered range of redshift is between $z_{\rm min}=5$ and
$z_{\rm max}=20$, because we are interested in the first
cosmic Gyr ($\sim\mbox{Hubble timescale at $z=5$}$) and the
number of
galaxies with $z_{\rm max}>20$ is negligible. The comoving
stellar mass density formed by $z=10$, 7, and 5
are $2.2\times 10^5~M_\odot~{\rm Mpc}^{-3}$,
$2.7\times 10^6~M_\odot~{\rm Mpc}^{-3}$, and
$1.3\times 10^7~M_\odot~{\rm Mpc}^{-3}$, respectively. In
units of the critical density, these numbers correspond to
$\Omega_\star =1.6\times 10^{-6}$,
$2.0\times 10^{-5}$, and $9.6\times 10^{-5}$, respectively.

As stated in \S~\ref{subsubsec:chemi}, the suppression of \H2
formation on grains for $T_{\rm dust}\ga 20$ K as shown
experimentally by Katz et al.\ (1999) will introduce a
characteristic star
formation epoch ($z_{\rm burst}$). Their results indicate
that \H2 formation on grains is allowed after the CMB
temperature, $T_{\rm CMB}$, drops to $\la 20$ K
($z\la 7$).
Since the \H2 formation on dust is crucial to activate
star formation in a galaxy, galaxies start to form stars
actively at $z_{\rm burst}\sim 7$ in this scenario.
Motivated by this, we have also investigated another star
formation history activated at $z\sim z_{\rm burst}$. We call
such a star formation history
``bursting star formation history'' (bSFH). On the other hand,
we call the star formation history without $z_{\rm burst}$
``continuous SFH'' (cSFH; the solid line of
Fig.\ \ref{fig:cosm_sfh}). Unless otherwise stated, we show the
results for the cSFH.

Galaxies with an active star formation are found up to
$z\sim 6.5$ (e.g., Hu et al.\ 2002). Moreover,
Shanks et al.\ (2001) have shown that the space density of
bright galaxies at $z\sim 6$ is comparable to the local space
density (see also Ouchi et al.\ 2002), which suggests that the
formation epoch of galaxies lies at $z>6.5$. In this paper,
we adopt the lower limit $z_{\rm burst}=6.5$. As shown later
on, even for this low $z_{\rm burst}$ value, ALMA and
{\it NGST} cannot clearly identify the existence of
$z_{\rm burst}$. However, it is worth examining the typical
flux level where the effect of $z_{\rm burst}$ on number
counts is clearly seen.

Even in the presence of the typical burst redshift, $z_{\rm burst}$,
the number of the virialised galaxies does not change. The only
difference is that the star formation is largely suppressed for
$z>z_{\rm burst}$ in the bSFH. Moreover, the star formation rate
is typically enhanced by a factor of 5 at $4t_{\rm cir}$ due to
the enhancement of molecular formation on accumulated dust.
Therefore, for $\Psi (z)$ in the bSFH, we adopt a value equal to
1/5 of the one calculated in equation (\ref{eq:cosmic_sfh})
if $z>z_{\rm burst}$. For $z<z_{\rm burst}$, we adopt the same
value both for cSFH and for bSFH.
The bSFH with $z_{\rm burst}=6.5$ is shown by the dashed line
in Fig.\ \ref{fig:cosm_sfh}.

\subsection{Galaxy number counts}\label{subsec:nc}

We estimate the number of galaxies with observed flux greater than
$f_\nu$ at an observed frequency $\nu$ by
\begin{eqnarray}
N(>f_\nu,\,\nu ) & = & \int_{z_{\rm min}}^{z_{\rm max}}dz\,
\int_{M_{\rm lim}(f_\nu ,\, z)}^{M_{\rm max}}dM \,\left[
\frac{\partial n(M,\, z)}{\partial M}\right.\nonumber\\
& \times & \left.\frac{dV(z)}{dz}\right]\, ,
\label{eq:nc_form}
\end{eqnarray}
where $M_{\rm lim}(f_\nu ,\, z)$ corresponds to the mass of the
galaxy with specific luminosity $L_\nu$, and
\begin{eqnarray}
f_\nu =\frac{(1+z)L_{\nu (1+z)}}{4\pi d_{\rm L}^2}\, .
\end{eqnarray}
The luminosity
distance, $d_{\rm L}$ is given by (Carroll, Press, \& Turner 1992)
\begin{eqnarray}
d_{\rm L}=c(1+z)\int_0^zdz'\, (1+z')\left|\frac{dt}{dz'}\right|\, ,
\end{eqnarray}
where $c$ is the light speed, and
\begin{eqnarray}
\left|\frac{dt}{dz}\right|^{-1} & = & H_0(1+z)\nonumber \\
& \times & \sqrt{(1+\Omega_{\rm M}z)(1+z)^2-\Omega_\Lambda z(2+z)}\, .
\end{eqnarray}
The volume element per unit redshift, $dV/dz$, is written as
\begin{eqnarray}
\frac{dV(z)}{dz}=\frac{4\pi cd_{\rm L}^2}{(1+z)}\left|
\frac{dt}{dz}\right|\, .
\end{eqnarray}
We assume that $L_\nu$ depends on $(M,\, z)$ and that
$L_\nu$ monotonically increases as $M$ for any $z$ and any $\nu$.
The monotonicity is guaranteed in the sub-mm number counts, but
not in the NIR counts. A NIR observation can detect the UV light
in the restframe of galaxies, and the UV luminosity is
not a monotonically increasing function of $\Mvir$ for
$\zvir > 15$ (Table \ref{tab:sfr}). Therefore, in calculating the
NIR number counts, we set $z_{\rm max}=15$. If
$M_{\rm lim}(f_\nu ,\, z)>M_{\rm max}$, we set
$M_{\rm lim}(f_\nu )$ equal to $M_{\rm max}$.

Finally, we must fix $L_\nu$ as a function of $M$ and $z$.
We approximately identify $(M,\, z)$ with $(\Mvir ,\,\zvir )$,
and use the quantities listed in Table \ref{tab:sfr} as a
function of $(\Mvir ,\,\zvir )$. The details of the calculation
will be described separately in the following two
subsubsections for sub-mm and NIR, respectively.

\begin{figure*}
\includegraphics[width=8cm]{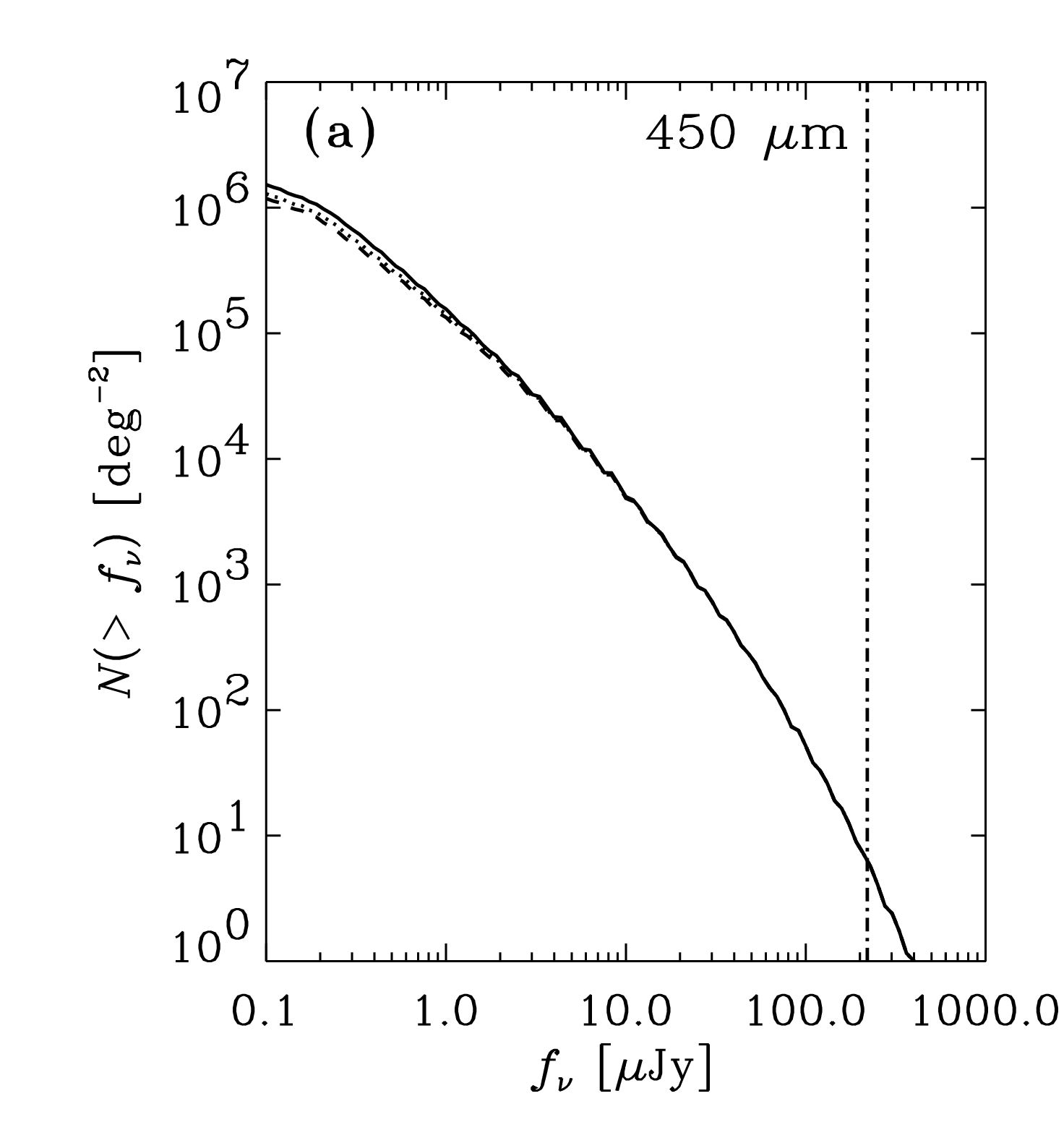}
\includegraphics[width=8cm]{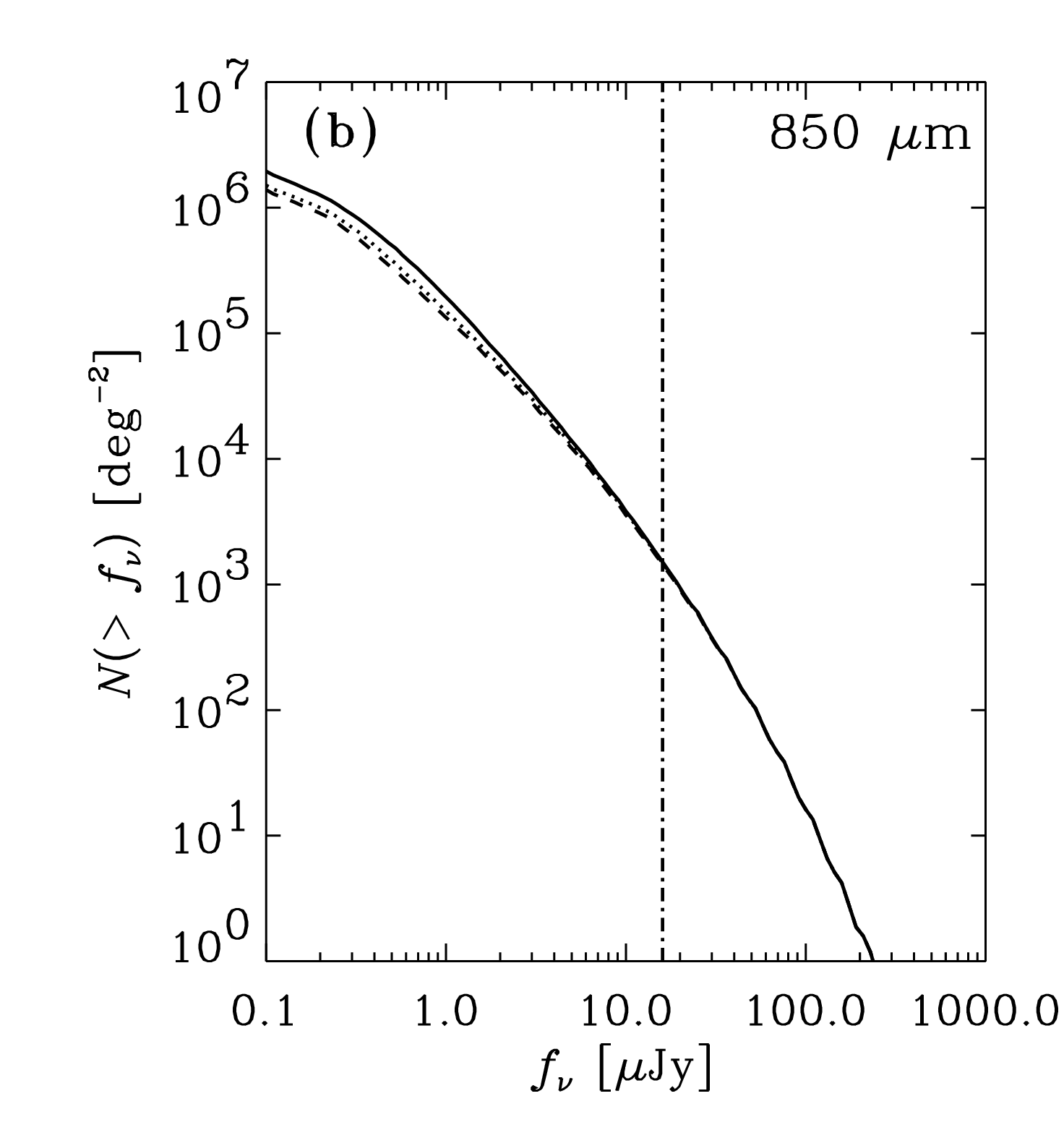}
\includegraphics[width=8cm]{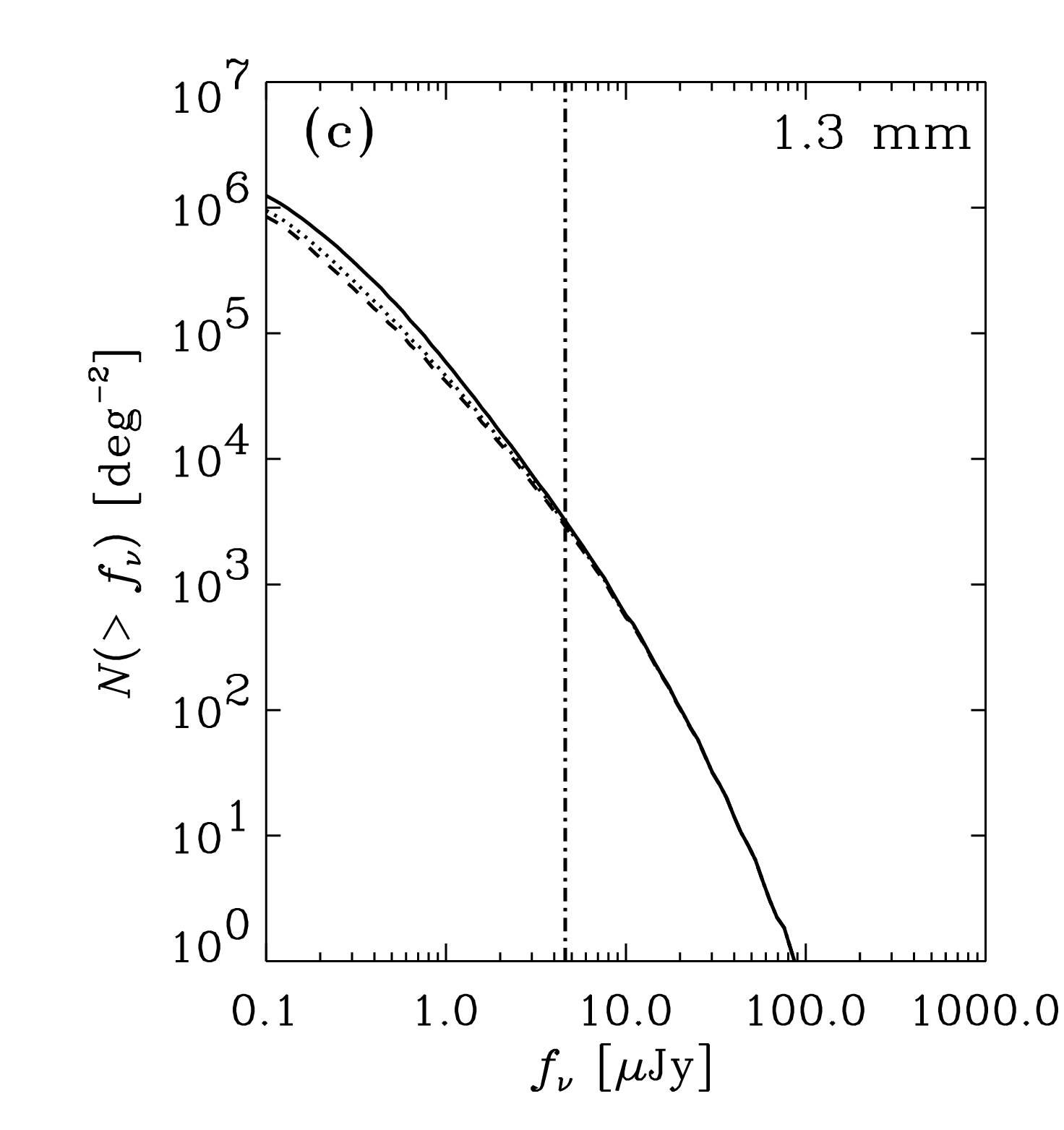}
\caption{Theoretical predictions of the number counts of the
high-redshift ($5<z<20$) galaxies in the ALMA bands (solid lines):
(a) 450 $\mu$m, (b) 850 $\mu$m, and (c) 1.3 mm.
The number counts of galaxies whose redshift range is $5<z<7$ are
also shown (dotted lines). The dashed lines show the result for
burst mode star formation (bSFH), in which galaxies start to
form stars actively at $z=6.5$. The vertical dot-dashed lines
show the ALMA detection limits.\label{fig:alma}}
\end{figure*}

\subsubsection{Submillimetric bands}\label{subsubsec:nc_submm}

We calculate the number counts due to continuum emission from
dust. The peak of dust emission falls in the FIR range
($\sim 50$--150 $\mu$m) in the restframe of a galaxy. The peak
of high-redshift ($z>5$) galaxies is shifted to the sub-mm range
in the observational restframe. For example, for a galaxy at
$z\sim 7$, observations around $\sim 800~\mu$m should be
performed. Here we consider three ALMA bands centered at 450,
850, and 1300 $\mu$m.

The spectral energy distribution of the FIR emission from a
galaxy is assumed to be a modified blackbody with a
temperature $\bar{T}_{\rm dust}$ (listed in Table \ref{tab:sfr}),
\begin{eqnarray}
L_\nu =C\nu^2B_\nu (\bar{T}_{\rm dust}) \, ,\label{eq:firsed}
\end{eqnarray}
where the coefficient $C$ is determined so that the integration of
$L_\nu$ for all the wavelength range becomes equal to
$\bar{L}_{\rm FIR}(\Mvir =M,\,\zvir =z)$.

{}From the above we can relate $M$ and $L_\nu$ for each $z$
using the relation between $L_\nu$ and
$(\bar{L}_{\rm FIR},\, \bar{T}_{\rm dust})$, and then
equation (\ref{eq:nc_form}) to obtain $N(f_\nu ,\,\nu )$. In
Figs.\ \ref{fig:alma}a--c, we show the number counts in three of
the ALMA observational bands of 450, 850, and 1300 $\mu$m
for galaxies with $5<z<20$ (solid lines) and with $5<z<7$
(dotted lines). The vertical lines show the detection limits,
for which we adopt
the same values as Takeuchi et al.\ (2001b) for the
5 $\sigma$ limits of 8 hr integration (220, 16, and
4.6 $\mu$Jy, respectively). We see that $5$, $1.5\times 10^3$,
and $2.9\times 10^3$ galaxies
per square degree can be detected in the 450, 850, and
1300 $\mu$m bands, respectively. The high angular resolution
of ALMA enables us to detect those galaxies without confusion.
A bright source can make the detection of faint sources
difficult because of a limited dynamic range of the detector.
The probability that sources which are $10^3$
(this number comes from the dynamic range of the detector of
ALMA) times larger than the 5 $\sigma$ detection limit of ALMA
exist in the field of view is negligible
($\la 3\times 10^{-4}$; Takeuchi et al.\ 2001b). From
Figure \ref{fig:alma}, it is also concluded that
ALMA is sensitive to galaxies with $z\la 7$.

\subsubsection{Near infrared bands}\label{subsubsec:nir_nc}

In order to test our theoretical prediction further,
it is necessary to detect the stellar light, because dust
production history is deeply related to
star formation history. Since young galaxies are
characterized by a strong UV stellar continuum produced by OB
stars, we examine the
detectability of the UV light, which is
redshifted to the NIR range in our observational restframe. For
example, an observation at 1.8 $\mu$m can detect the stellar
light at 3000 \AA\ in the restframe of a galaxy at $z=5$, or
that at 2000 \AA\ at $z=8$. In order to calculate the NIR number
counts, we must fix the UV spectra. Although we have not
included spectral synthesis model of stellar populations (but
we have used the stellar model by Schaerer 2002 in
\S~\ref{subsubsec:UV_FIR}), the following constant spectral energy
distribution can be used as a first approximation to the NIR
number counts:
\begin{eqnarray}
L_\nu (M,\, z)=\bar{L}_{\rm UV}(\Mvir =M,\,\zvir =z)/\Delta
\nu_{\rm UV}\, ,
\end{eqnarray}
where $\Delta\nu_{\rm UV}$ is the typical width of the
frequency range ($\sim 10^{15}$--$8\times 10^{15}$ Hz;
400--3000 \AA\ in wavelength) where OB stars dominate the
luminosity. Here we assume
$\Delta\nu_{\rm UV}=7\times 10^{15}$ Hz. We use the data listed
in Table \ref{tab:sfr} for $\bar{L}_{\rm UV}$ as a function of
$(\Mvir ,\,\zvir )$.

\begin{figure}
\includegraphics[width=8.5cm]{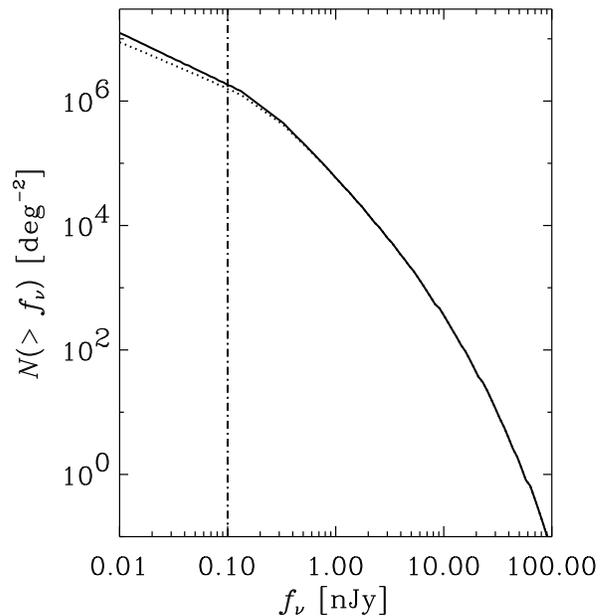}
\caption{Number counts for high-redshift galaxies in
the near infrared. The solid and dotted lines show the number
counts in the redshift range of $5<z<20$ and $5<z<7$,
respectively. The vertical dotted line indicates an expected
detection limit (0.1 nJy) of the {\it NGST} photometry
(Gardner \& Satyapal 2000).\label{fig:ngst}}
\end{figure}

In Fig.\ \ref{fig:ngst}, we show the number counts in NIR. The
solid and dotted lines show the counts for $5<z<20$ and
$5<z<7$, respectively. The vertical dot-dashed line shows a
proposed detection limit of {\it NGST} at the NIR
($\simeq 0.1$ nJy; Gardner \& Satyapal 2000). We see that about
$2\times 10^6$ galaxies should be detected per square degree.
The detection limit can be larger (1 nJy;
Mather \& Stockman 2000). In this case, about $10^5$ galaxies
are detected per square degree. These numbers are much larger
than those derived for ALMA. In other words, all the ALMA
sample for $z>5$ is detected by {\it NGST}. Therefore,
{\it NGST} can be used for the two aims: for an identification of
ALMA sources and for an exploration of the universe deeper than ALMA.
{\it NGST} can also detect galaxies whose redshift is larger than 7.
The high-redshift $z>5$ sample brighter than 0.1 nJy (1 nJy)
contains 14\% (1.2\%) of ``extremely'' high-redshift $z>7$ sources.

\subsubsection{bSFH vs. cSFH in sub-mm number counts}

Here, we consider the bSFH. The difference between bSFH
and cSFH is larger in sub-mm number counts than in
NIR number counts, because dust accumulation makes the
FIR luminosity increase efficiently while UV suffers from
extinction by dust. The FIR luminosity increases by
an order of magnitude owing to the increase of
star formation rate and dust content. Therefore,
the effect of the bSFH can be
examined by adopting a FIR luminosity 1/10 times smaller than for
cSFH for $z>z_{\rm burst}$. For $z>z_{\rm burst}$, we adopt the
same star formation rate as that of cSFH. As a
result, the FIR luminosity rises by 10 times
at $z=z_{\rm burst}$ in the bSFH.
The expected sub-mm number counts for the bSFH is shown by
the dashed lines in Figure \ref{fig:alma}. A clear difference
between the two scenarios cannot be seen at the ALMA
detection limit. Since ALMA is not sensitive
to galaxies at $z\la 7$ (\S~\ref{subsubsec:nc_submm}),
we cannot distinguish the two SFHs as long as the difference
appears at $z\la 7$.

\subsection{Integrated light of UV and FIR emissions}

We also predict the level of the flux integrated for
all the galaxies from $z_{\rm max}=20$ to $z_{\rm min}=5$,
to see whether our model is consistent with the current
observational constraints, and to compare our result with future
more sensitive observations. The integrated light from all the
extragalactic objects (extragalactic background radiation) has
been detected in a wide range of wavelength (e.g., Hauser \& Dwek
2001). We can examine what fraction of the extragalactic
background light is produced by the high-redshift galaxies by
using our results.

Theoretically, the intensity of integrated light produced by
sources between $z_{\rm min}$ and $z_{\rm max}$ at an observed
frequency $\nu$ is estimated by
\begin{eqnarray}
I_\nu & = & \int_{z_{\rm min}}^{z_{\rm max}}dz
\int_{M_{\rm min}}^{M_{\rm max}}dM\left[
\frac{\partial n(M,\, z)}{\partial M}\,\frac{dV(z)}{dz}\right.
\nonumber \\
& \times & \left.\frac{(1+z)L_{(1+z)\nu}}{4\pi d_{\rm L}^2}\right]\, .
\end{eqnarray}
The same spectra ($L_\nu$) as \S~\ref{subsec:nc} are adopted for
both FIR and UV.

\begin{figure}
\includegraphics[width=8.5cm]{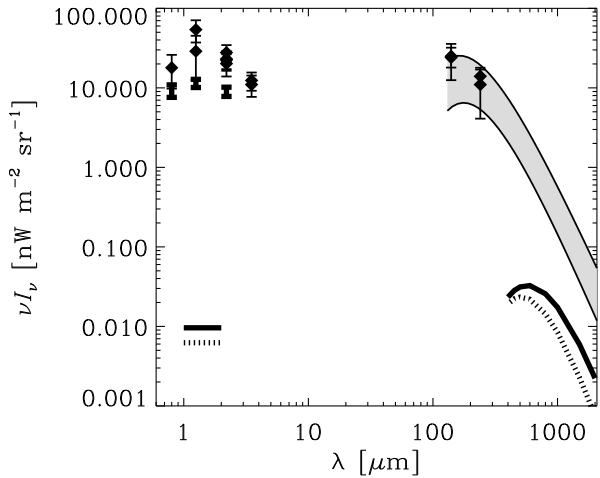}
\caption{Predicted intensity levels of integrated light from all
the high-redshift ($5<z<20$) galaxies for (sub-)millimetre and
near-infrared wavelengths (thick solid lines). The contribution
from galaxies with $5<z<7$ is also calculated (thick dotted 
lines). We have also calculated the integrated light from all
the high-redshift galaxies in the bSFH scenario, but the results
are indistinguishable from the thick dotted lines in both
UV and (sub-)millimetre. The observational data for
extragalactic background radiation are also shown because
our predictions should lie below them. The references are as
follows: Bernstein et al.\ (2002; 0.8 $\mu$m);
Cambr\'{e}sy et al.\ (2001; 1.25 and 2.2 $\mu$m);
Wright (2001; 1.25 and 2.2 $\mu$m);
Gorjian et al.\ (2000; 2.2 and 3.5 $\mu$m);
Wright \& Reese (2000; 2.2 and 3.5  $\mu$m);
Hauser et al.\ (1998; 140 and 240 $\mu$m);
Lagache et al.\ (2000; 140 and 240 $\mu$m).
We also show the ranges estimated from integration of NIR galaxy
counts by Totani et al.\ (2001; thick error bars without
symbols). The shaded area indicates the observed level
constrained by {\it COBE} measurement (extragalactic background;
Fixsen et al.\ 1998).
\label{fig:bacg_submm}}
\end{figure}

The sub-mm and NIR integrated intensities of the high-redshift
galaxies are
shown in Fig.\ \ref{fig:bacg_submm} (thick solid lines). Both
fluxes are well below the observed extragalactic background
radiation. The largest ($\simeq 10$\%) contribution of the
high-redshift galaxies  is expected in the millimetre range.
This indicates that
future ALMA 1.3 mm observations will be the most efficient
to isolate the high-redshift contribution.

In Figure \ref{fig:bacg_submm}, we
also show the integrated intensities in the case of
$z_{\rm max}=7$ and $z_{\rm min}=5$ (thick dotted lines).
Comparing thick solid and dotted lines, we see that roughly
52\% and 35\% of the high-redshift ($z>5$) integrated light
comes from sources with $z>7$ at millimetre and UV
wavelengths, respectively.
The model predictions in the case of the bSFH are also shown
but the results are indistinguishable from the thick dotted
lines (the difference is at most 7\%). Therefore, if the
bSFH is correct, the integrated light
from $z>5$ sources is reduced by half in the millimetre range.

For NIR, a more elaborate model focusing on Pop III stars
has been developed by
Santos, Bromm, \& Kamionkowski (2002) and Salvaterra \& Ferrara
(2002). These authors took into account the formation of Pop III
stars in halos less massive than our $M_{\rm min}$. They also
considered stars whose mass is
larger than $100~M_\odot$ because
Pop III stars are widely believed to be massive. Our model, on
the other hand, has included
$M_{\rm min}$ to exclude galaxies that cannot sustain gas
against energy input from SNe II, but adopted a more standard
IMF. The integrated light of NIR by Santos et al.\ (2002) is two
orders of magnitude higher than our integrated light
of high-redshift galaxies, although we should note that
``extreme'' conditions (star formation efficiency of 40\%,
activation of star formation in halos less massive than
$M_{\rm min}$) are required in their paper to realise such a
high value of integrated flux.

\subsection{Integrated metal-line intensity}
\label{subsec:integ_metal}

Suginohara et al.\ (1999) have proposed that the spatial
fluctuations of integrated metal-line intensity can be used as
a tracer of
metal production in high-redshift sources. Now that we have
analysed the metal production and the star formation history
in a consistent manner, it is worth reexamining the integrated
metal-line intensity within our framework.

We consider an observation which is sensitive to the
frequency range
$[\nu_{\rm obs}-\Delta\nu_{\rm obs}/2 ,\,
\nu_{\rm obs}+\Delta\nu_{\rm obs}/2]$. If we are to detect a
metal line whose frequency is $\nu_{\rm line}$, the redshift
range which we can observe is $[z_1,\, z_2]\equiv
[\nu_{\rm line}/(\nu_{\rm obs}+\Delta\nu_{\rm obs}/2)-1,\,
\nu_{\rm line}/(\nu_{\rm obs}-\Delta\nu_{\rm obs}/2)-1]$.
Therefore, the integrated line intensity,
$I_{\rm line}(\nu_{\rm obs};\,\Delta\nu_{\rm obs})$, is
\begin{eqnarray}
I_{\rm line}(\nu_{\rm obs};\,\Delta\nu_{\rm obs}) & = &
\int_{z_1}^{z_2}dz\,\int_{M_{\rm min}}^{M_{\rm max}}dM\left[
\frac{\partial n(M,\, z)}{\partial M}\right.
\nonumber \\
& \times & \left.\frac{dV(z)}{dz}
\frac{L_{\rm line}}{4\pi d_{\rm L}^2}\right]\, ,
\end{eqnarray}
where we assume that $\Delta\nu$ is much wider than
the line width of each galaxy. In this paper, we assume
$z_1=5$ and $z_2=7$ to concentrate on high-redshift
galaxies detected by ALMA (\S~\ref{subsubsec:nc_submm}).
Here, we again approximate $(M,\, z)$ with
$(\Mvir ,\,\zvir )$. In Table \ref{tab:metal}, we show
$I_{\rm line}^{\rm max}$ relative to the CMB intensity,
where the superscript ``max'' indicates that the integrated
line intensity is calculated by using the maximum line
intensity as listed in Table \ref{tab:sfr}.
We also present $\nu_{\rm obs}I_{\rm line}^{\rm max}$
in Table \ref{tab:metal}.

\begin{table*}
\begin{center}
\caption{Metal lines considered in this paper}
\begin{tabular}{lccccccc}\hline
Species & $m_i$  & transition & Wavelength & $A_{\rm ul}$ &
$n_{\rm H}^{\rm c}$ & $I_{\rm metal}^{\rm max}/I_{\rm CMB}$
&$\nu_{\rm obs} I_{\rm metal}^{\rm max}$ \\
& ($M_\odot$) & & ($\mu$m) & (s$^{-1}$) & (cm$^{-3}$) & &
(nW m$^{-2}$)
\\ \hline
C & 0.17 & $^3P_1\to{}^3P_0$ & 609 & $7.93\times 10^{-8}$ &
           $4.7\times 10^2$ & $3.6\times 10^{-5}$ &
           $4.3\times 10^{-4}$ \\
  &      & $^3P_2\to{}^3P_2$ & 370 & $2.68\times 10^{-7}$ &
           $2.8\times 10^3$ & $1.5\times 10^{-6}$ &
           $4.6\times 10^{-5}$ \\
O & 1.2  & $^3P_1\to{}^3P_2$ & 63.2 & $8.95\times 10^{-5}$ &
           $4.7\times 10^5$ & 11 & 13 \\
  &      & $^3P_0\to{}^3P_1$ & 146 & $1.70\times 10^{-5}$ &
           $9.5\times 10^4$ & $7.2\times 10^{-3}$ & 0.33 \\
C$^+$ & 0.17 & $^2P_{1/2}\to{}^2P_{3/2}$ & 158 & $2.36\times 10^{-6}$ &
           $2.8\times 10^3$ & $1.0\times 10^{-4}$ &
           $5.3\times 10^{-3}$ \\ \hline
\end{tabular}
\label{tab:metal}
\end{center}
\end{table*}

As discussed in \S~\ref{subsubsec:metal}, the metal-line
intensity should be multiplied by ${\cal F}<1$ if the gas
density of
galaxies is lower than $n_{\rm H}^{\rm c}$. Adopting the
mean density in \S~\ref{subsubsec:density}, we obtain
${\cal F}\simeq 5.5\times 10^{-2}$, $9.7\times 10^{-3}$,
$5.8\times 10^{-5}$, $2.9\times 10^{-4}$, and
$9.7\times 10^{-3}$ at $z\sim 6$ (average between $z_1$ and
$z_2$) for C {\sc i} 609 $\mu$m,
C {\sc i} 370 $\mu$m,
O {\sc i} 63.2 $\mu$m, O {\sc i} 146 $\mu$m, 
and C {\sc ii} 158 $\mu$m, respectively. Hence intensities
are reduced by the small ${\cal F}$ values and are difficult
to detect. However, if we adopt, for example, the
typical density for photo-dissociation region in the nearby
universe $\sim 10^3~{\rm cm}^{-3}$ (Hollenbach \& McKee 1979),
we obtain ${\cal F}\sim 0.68$, 0.26, $2.1\times 10^{-3}$,
$1.0\times 10^{-2}$, and 0.26, for C {\sc i} 609 $\mu$m,
C {\sc i} 370 $\mu$m,
O {\sc i} 63.2 $\mu$m, O {\sc i} 146 $\mu$m, 
and C {\sc ii} 158 $\mu$m, respectively.
In Figure \ref{fig:metal_dilution}, we show ${\cal F}$ as a
function of hydrogen number density. Observations of the
integrated line intensity will constrain
the typical gas density in high-redshift galaxies by 
comparison with theoretical maximum values listed in
Table \ref{tab:metal}. If more than two metal lines are
detected, a consistency check of gas density is possible.

\section{SUMMARY AND DISCUSSIONS}\label{sec:discussion}

\subsection{Summary of evolutionary properties}

In order to quantify the importance of dust on the first
star formation activity in the universe, we  have solved
the time evolution of dust mass in the galaxies formed
in the redshift range $z>5$, when the age of the universe is
$\la 1$ Gyr. We have taken into account the importance of
\H2 abundance for the star formation rate, and the
formation of molecules on dust in a consistent manner
(\S~\ref{sec:each}). In particular, we have made the
first attempt to tie the star formation efficiency to \H2
abundance in relatively primordial environments
(\S~\ref{subsec:sfr}). Even when this inefficient phase of
star formation is included, an
active phase of star formation takes place after a few
$t_{\rm cir}$ (much shorter than the Hubble timescale) because a
significant amount of dust is accumulated to activate the \H2
formation on the grain surfaces (Fig.\ \ref{fig:H2}). This
suggests that the grains play
an essential role in causing the first active phase of
star formation. As a result, we have provided a robust support
for some theoretical works that have implicitly assumed that
stars are formed at high redshift as efficiently as at low
redshift.

Radiative properties of high-redshift star-forming galaxies are
also predicted. We have found that a significant amount of
luminosity is radiated in the FIR range. The FIR luminosity
becomes comparable
to the UV luminosity in a few $t_{\rm cir}$, when a significant
amount of dust is accumulated (Fig.\ \ref{fig:fir_luminosity}).
This efficient reprocessing of 
the UV light into the FIR results partly from the dense
(i.e., large optical depth) environment of high-redshift
galaxies. Observations
of the FIR light (sub-mm light in the observer's restframe) as
well as those of the stellar light are thus crucial to
trace the whole stellar radiative energy from high-redshift
galaxies.

In the framework of our model we have also given an approximate
estimate of some sub-mm metal-line luminosities. However, this
can only been seen as an upper limit to the actual luminosity
because considerable
uncertainty is present on the gas density
(\S~\ref{subsubsec:metal} and Appendix \ref{app:line}). If
future sub-mm or millimetre observations detect metal lines, a
density probe of ISM of the high-redshift galaxies
will be possible. If more than two types of metal-lines are
detected, density can be estimated more precisely.

\subsection{Future observational tests}

In about ten years, it will become possible to detect sources at
high redshift ($z>5$) in both sub-mm and NIR. The luminosity
level of the galaxies at these wavelengths will put important
constraints on our model. Therefore, we have calculated the
number counts for both wavelengths. As a result, we have found
that ALMA (450 $\mu$m, 850 $\mu$m, and 1.3 mm bands) and
{\it NGST} (NIR bands) can detect
several times $10^3$ and $10^6$ high-redshift galaxies per square
degree, respectively. These numbers can be used to test our model.

We have also calculated the integrated intensity of the metal-line
emission from the galaxies from $z=5$ to 7. Although a precise
determination of the line intensity requires a model for the
gas density (perhaps a model for photo-dissociation region as
Hollenbach \& McKee 1979 is also necessary), we can estimate a
maximum intensity for the
integrated metal-line intensity. The results are listed in
the last two columns of Table \ref{tab:metal}. The contamination
with the cosmic sub-mm and microwave background could represent
a potential
problem, but such high-redshift galaxies have a correlation
scale of the order of $10''$ (Appendix \ref{app:angular}).
Therefore, a fluctuation analysis of sky brightness in
(sub-)millimetre can discriminate the metal-line signal from
other contaminating sources by examining the typical correlation
scale. A quantitative analysis of the fluctuations using the
structure formation theory is left for
future work. It may be also possible to probe the gas density of
high-redshift galaxies through the factor ${\cal F}$ in
Appendix \ref{app:line}, if more than two kinds of metal-lines
are detected. The relative intensities of two lines can be
used to derive a probable value of $\nH$.

\begin{figure}
\includegraphics[width=8cm]{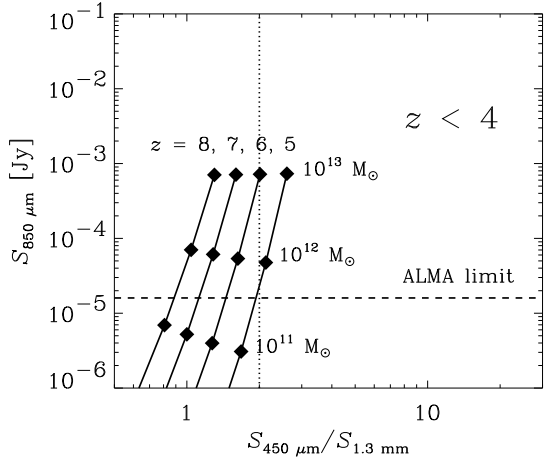}
\caption{Relation between the flux at $\lambda =850~\mu$m
($S_{\rm 850~\mu m}$) and the 450--1300 $\mu$m color
($S_{\rm 450~\mu m}/S_{\rm 1.3~mm}$). (We adopt the same notations
as Fig.\ 10 of Takeuchi et al.\ 2001b in this figure.) Our model
predictions are identified by filled squares for $z=5$, 6, 7,
and 8 and $\Mvir =10^{11}$,
$10^{12}$ and $10^{13}~M_\odot$. The horizontal dashed line
indicates the detection limit of ALMA. Low-redshift galaxies
($z<4$) typically fall on the right of the vertical dotted line
(Takeuchi et al.\ 2001b).
\label{fig:color}}
\end{figure}

Finally, we should mention how to select high-redshift galaxies
efficiently. Galaxy colors (flux ratios between two bands) are
often used for the selection. We have shown that ALMA can
detect galaxies at $z\la 7$. The peak of the dust emission
lies roughly between 500 $\mu$m and 800 $\mu$m for galaxies
between $z=5$ and 7. For example, 450 $\mu$m vs. 1.3 mm flux
ratio (450--1300 $\mu$m color) gives us a useful information on
the redshift, because the peak of flux lies between these bands
only for the high redshift galaxies. In Figure \ref{fig:color},
we show the flux at 850 $\mu$m ($S_{850~\mu{\rm m}}$) and
450--1300 $\mu$m color ($S_{450~\mu{\rm m}}/S_{1.3~{\rm mm}}$)
predicted by the modified blackbody spectra
(eq.\ \ref{eq:firsed}).
The galaxies detected by ALMA in that redshift range has typical
flux levels $\sim 10$--100 $\mu$Jy and the 450--1300 $\mu$m
color is typically less than 3. In Figure \ref{fig:color}, we
also show the ALMA detection limit
(horizontal dashed line). Galaxies with
$\Mvir\ga 10^{11.5}~M_\odot$ will be detected by ALMA.
At $z>7$, however, the number of such massive galaxies is
negligible and does not contribute to number counts.
Typically, galaxies whose redshift is less than 4 fall to the
right of the vertical dotted line (Takeuchi et al.\ 2001b).

\begin{figure}
\includegraphics[width=8cm]{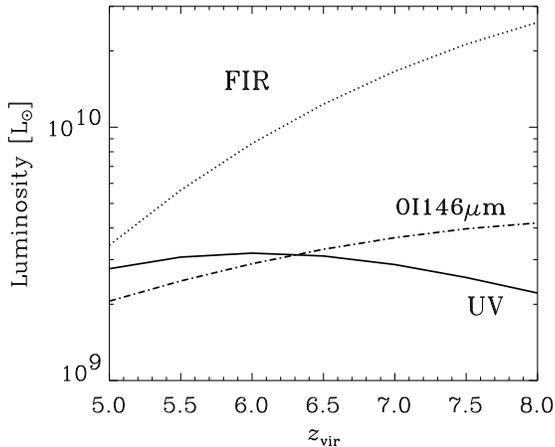}
\caption{Luminosities as a function of formation redshift
$\zvir$ for $\Mvir =10^{11.5}~M_\odot$. Such a massive
galaxy will be detected by ALMA.
The solid, dotted, and dashed lines represent ultraviolet,
far-infrared, and O {\sc i} 146 $\mu$m (maximum;
$\bar{L}_{\rm O146}^{\rm max}$) luminosities defined
at 4 circular times.
\label{fig:lum_vs_z}}
\end{figure}

We can also select high-redshift galaxies efficiently from
optical observations by using the ``dropout'' technique
(Steidel et al.\ 1996). The Lyman limit at the wavelength of
912 \AA\ in the restframe of a galaxy is redshifted to
5500--7300 \AA\ for galaxies at $z=5$--7. Therefore,
optical/NIR observations of galaxies by {\it NGST}
(Mather \& Stockman 2000) or other sensitive facilities provides
us a way to sample the
high-redshift candidates independent from the ALMA
sample. A large sample of galaxies with $V$- or $R$-band
dropout should be collected by future observations. After spatial
cross identification of drop-out sample with ALMA sample,
we can investigate the optical--sub-mm flux ratio
as a test of our model. Galaxies with
$\Mvir\ga 10^{11.5}~M_\odot$ are detectable both by ALMA and
{\it NGST}. In order to see the typical luminosities for
galaxies detected by
ALMA, we show in Figure \ref{fig:lum_vs_z} $\bar{L}_{\rm UV}$
and $\bar{L}_{\rm FIR}$ as a function of $\zvir$. We see that
FIR/UV flux ratios are 1.2, 2.7, and
5.8 for $\zvir =5$, 6, and 7, respectively. In the same
figure, we also present $\bar{L}_{\rm O146}^{\rm max}$.
As mentioned in \S~\ref{subsec:integ_metal}, the ratio
between the observed line luminosity and
$\bar{L}_{\rm O146}^{\rm max}$ (i.e., ${\cal F}$) can
be used to estimate the density of ISM.

\subsection{Connection to lower redshift}

As shown in Figure \ref{fig:cosm_sfh}, our predictions connect
smoothly to the lower-redshift star formation history. Our model,
however, cannot be applied to galaxy evolution at
$z<5$ because after that epoch dust is supplied from late-type
stars as well as SNe II. It is
observationally known that mergers between
giant galaxies significantly contribute to
luminous infrared populations at the local universe
(Sanders \& Mirabel 1996) and even at $z\sim 1$
(Roche \& Eales 1999). When we apply our framework to lower
redshifts, therefore, it is necessary to extend our model to
include the details of the merging history of galaxies. We
should note that the
enhancement of molecular formation is also
a key to star formation activity in mergers (e.g.,
Walter et al.\ 2002).

Recent studies using the Subaru telescope (Ouchi et al.\ 2002)
have pushed observations as deep as $z\sim 5$. Therefore, the
luminosity function (or comoving star formation rate) derived
from the ``Subaru
Deep Field'', which is as wide as 600 arcmin$^2$
and as deep as 26 AB magnitude around 7000 \AA will allow us
to directly compare our results at $z\sim 5$. The luminosity
function at $z\sim 5$ is also important to constrain the
evolutionary scenario of Lyman break
populations found at $z\sim 3$. Is the luminosity function
of galaxies at $z\sim 5$ explained by the same
population of Lyman break galaxies at $z\sim 3$? Recently,
Ferguson, Dickinson, \& Papovich (2002) have given a
negative answer to this question, but further studies are
necessary to reveal the link between these two epochs.

We have stressed the importance of dust on the formation of
molecular-rich environment. In the lower-redshift ($z<5$)
universe, it is observationally known that there is a
correlation between the abundances of dust and molecules
for DLAs (Ge et al.\ 2001). This strongly suggests the
important role of dust for molecule formation (see also
Levshakov et al.\ 2002). However, the correlation is not
firmly assessed and further
observational sample seems to be required
(Petitjean et al.\ 2000). Petitjean et al.\ also noted that
most of DLAs may arise
selectively in warm and diffuse neutral gas. Liszt (2002) has
shown that even in a cool medium \H2 formation can be
suppressed because of low dust content and strong UV radiation
field. Moreover, DLA trace a diverse population with various
mass, surface
brightness, etc.\ (e.g., Pettini 2002). In spite of those
complexities, DLAs are promising objects to study the link
between the abundances of dust and
molecules in the early universe.

\section*{Acknowledgments}
We thank M. Edmunds, the referee, for helpful comments that
improved this paper very much. We also thank B. Ciardi,
T. T. Takeuchi, and M. Ouchi for useful
discussions and suggestions about various topics on galaxy
evolution. Some of our cosmological results
were checked against the codes provided by K. Yoshikawa.
We are grateful to T. T. Ishii for helping with IDL
programming. H. H. was supported by JSPS Postdoctoral
Fellowship for Research Abroad. We fully utilized the
NASA's Astrophysics Data System Abstract Service (ADS).

\appendix

\section{H$_2$ formation on grains}\label{app:surface_rate}

The production rate of molecular fraction via dust surface reaction
is estimated as (Hollenbach \& McKee 1979)
\begin{eqnarray}
\left[\frac{df_{\rm H_2}}{dt}\right]_{\rm dust}=f_0n_{\rm dust}
\pi a^2\bar{v}S\, ,\label{eq:H2_on_grain}
\end{eqnarray}
where $\fH2$ is the molecular fraction of hydrogen
(equation \ref{eq:mol_frac}), $f_0$ is the neutral fraction of
hydrogen, $\nH$ is the number density of hydrogen nuclei, $\bar{v}$
is the mean thermal speed of hydrogen, $n_{\rm dust}$
is the number density of dust grains, $a$ is the radius of a
grain (spherical grains with a single radius are assumed),
and $S$ is the sticking efficiency of hydrogen atoms. The
thermal speed is given by (Spitzer 1978)
\begin{eqnarray}
\bar{v}=\sqrt{\frac{8}{\pi}\frac{k_{\rm B}T}{m_{\rm H}}}=1.4
\times 10^5\left(\frac{T}{100~{\rm K}}\right)^{1/2}~{\rm cm~s}^{-1}
\, ,
\end{eqnarray}
where $k_{\rm B}$ is the Boltzmann constant, 
$T$ is the gas temperature, and $m_{\rm H}$ is
the mass of a hydrogen atom.
Here, we define the reaction rate of the \H2 formation on
grains, $R_{\rm dust}$, as
\begin{eqnarray}
R_{\rm dust} & \equiv & \frac{3m_{\rm H}\bar{v}S}{8a\delta}
\nonumber \\
 & = & 1.4\times 10^{-14}S\left(\frac{T}{100~{\rm K}}\right)^{1/2}
 \left(
 \frac{a}{0.03~\mu{\rm m}}\right)^{-1} \nonumber \\
 & \times & \left(\frac{\delta}{2~{\rm g~cm^{-3}}}
 \right)^{-1}~{\rm cm^3~s^{-1}}\, ,\label{eq:surface_rate}
\end{eqnarray}
where $\delta$ is the mass density of a grain. The dust-to-gas
mass ratio ${\cal D}$ can be estimated as
\begin{eqnarray}
{\cal D}=\frac{4\pi a^3\delta\, n_{\rm dust}}{3\nH m_{\rm H}}\, ,
\end{eqnarray}
Using $R_{\rm dust}$ and ${\cal D}$,
the right-hand side of equation (\ref{eq:H2_on_grain}) becomes
the second term of equation (\ref{eq:H2_form_rate}).
We take $a=0.03$ $\mu$m (Todini \& Ferrara 2001).
The sticking
coefficient $S$ is uncertain. Simply, we adopt
$S\sim 0.2$, the value for the gas temperature when star
formation occurs
($T<300$ K) (Hollenbach \& McKee 1979). When $T>300$ K, $S$ is
assumed to be zero. The sticking efficiency of
Hollenbach \& McKee (1979) indicates that the dependence of $S$
on dust temperature
is negligible as long as the dust temperature, $T_{\rm dust}$,
is less than $\sim 75$ K.

\section{Simple prescription to include the low-density effect on
line intensity}\label{app:line}

Here we describe a simple approximate treatment to calculate
FIR--sub-mm metal-line intensities. Our formula given here is
appropriate for a one-zone treatment as our model. A more
accurate treatment requires detailed modeling of
photo-dissociation regions (Hollenbach \& McKee 1979;
Tielens \& Hollenbach 1985).

We consider a population with two energy levels (the upper and
the lower are labelled as $u$ and $l$, respectively). We consider
spontaneous transition, whose Einstein coefficient
is denoted as $A_{ul}$, and collisional excitation by
species whose number density is $n$ (the collisional
excitation rate is expressed as $\gamma_{lu}n$) and
collisional deexcitation by the same species (the collisional
deexcitation rate is expressed as $\gamma_{ul}n$). Assuming
the equilibrium between the transition from $u$ to $l$ and
that from $l$ to $u$, 
the fraction of the population in the upper level, $f_u$, is
estimated to be
\begin{eqnarray}
f_u=\frac{\gamma_{lu}n}{A_{ul}+(\gamma_{ul}+\gamma_{lu})n}\, .
\end{eqnarray}
If the collisional excitation (or deexcitation) occurs on
shorter timescale than the spontaneous emission,
the fraction in the upper level is
\begin{eqnarray}
f_u^0=\frac{\gamma_{lu}}{\gamma_{ul}+\gamma_{lu}}\, .
\end{eqnarray}
Because of the spontaneous emission, the fraction in the
upper level is reduced by the factor of ${\cal F}\equiv
f_u/f_u^0$.
For the excitation coefficients, the
following relation holds (Spitzer 1978):
\begin{eqnarray}
\frac{\gamma_{lu}}{\gamma_{ul}}=\frac{g_u}{g_l}\exp\left( -
\frac{E_{ul}}{k_{\rm B}T_{\rm ex}}\right)\, ,
\end{eqnarray}
where $g_u$ and $g_l$ are statistical weights for the lower
and the upper levels, respectively, $E_{ul}$ is the energy
gap between the two states, $k_{\rm B}$ is the Boltzmann
constant, and $T_{\rm ex}$ is the excitation temperature
of the colliding species. In this paper, we consider metal
lines whose $\gamma_{lu}/\gamma_{ul}$ is order unity
(i.e., $E_{ul}\la k_{\rm B}T_{\rm ex}$) and whose exciting
species is hydrogen ($n\sim\nH$). For the metal lines of
interest, we finally find that
\begin{eqnarray}
{\cal F}\sim \frac{\nH}{n_{\rm H}^{\rm c}+\nH}\, ,
\label{eq:simplef}
\end{eqnarray}
gives a good approximation (correct within a factor of $\sim 2$),
where $n_{\rm cr}$ is the critical density defined as
\begin{eqnarray}
n_{\rm H}^{\rm c}\equiv\frac{A_{ul}}{\gamma_{ul}}\, .
\end{eqnarray}
The line intensity is expected to be $\sim L_{\rm line}^{\rm max}$
(\S~\ref{subsubsec:metal}) if $\nH$ is higher than
$n_{\rm H}^{\rm c}$ (i.e., ${\cal F}\sim 1$).
On the contrary, if the density if much smaller than
$n_{\rm H}^{\rm c}$, the line intensity is quite reduced
(i.e., ${\cal F}\ll 1$).

\begin{figure}
\includegraphics[width=8cm]{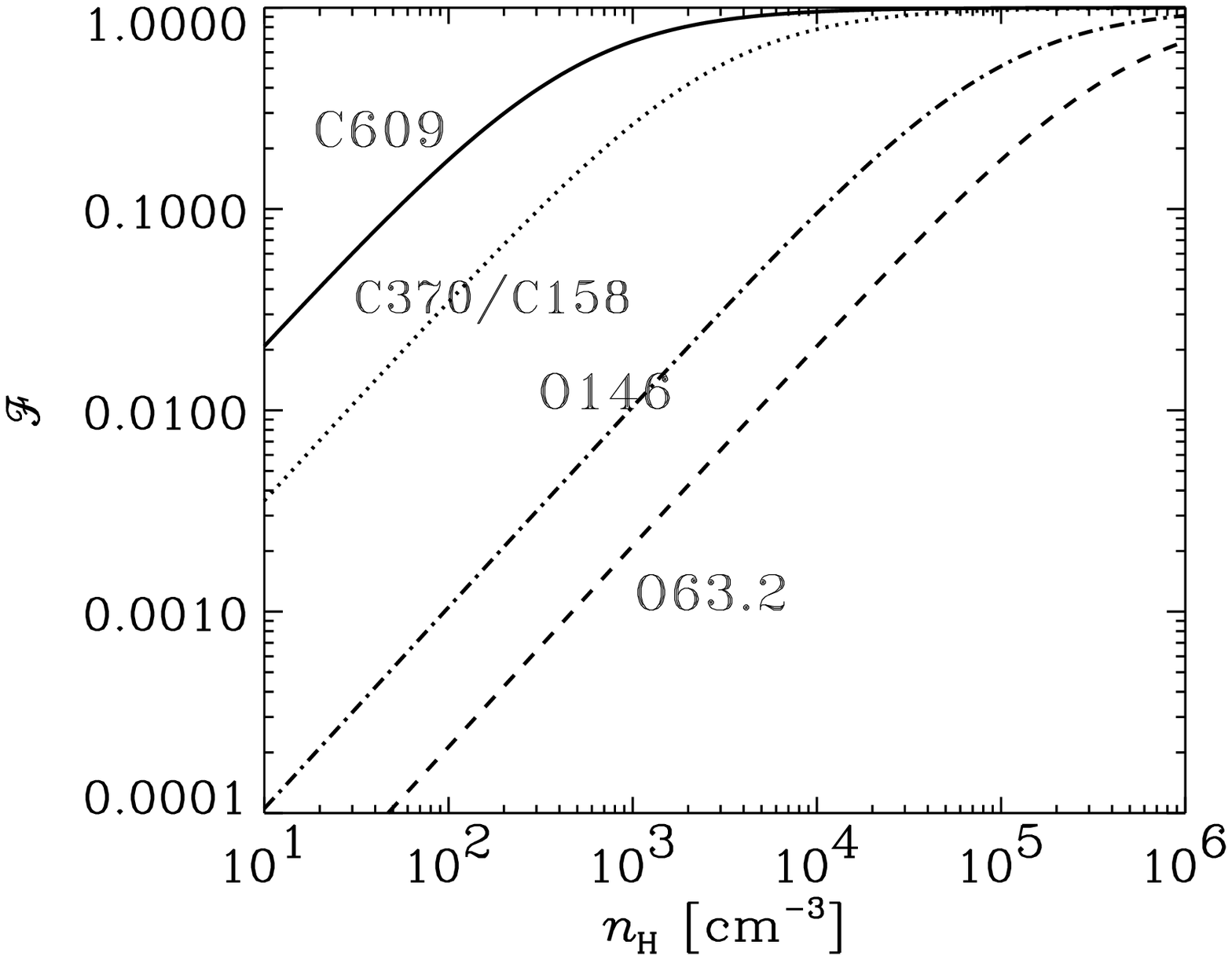}
\caption{Reduction factor ${\cal F}$ to the maximum luminosity
$L_{\rm line}^{\rm max}$ of metal lines as a function of
the number density of hydrogen atoms, $\nH$. The solid, dotted,
dashed, and dot-dashed lines represents ${\cal F}$ of
C {\sc i} 609 $\mu$m, C {\sc i} 370 $\mu$m,
O {\sc i} 63.2 $\mu$m, and
O {\sc i} 146 $\mu$m, respectively. Since the critical
density of C {\sc ii} 158 $\mu$m is the same as that of
C {\sc i} 370 $\mu$m, ${\cal F}$ of C {\sc ii} 158 $\mu$m
is represented by the dotted line.
If the number density $\nH$ is known, we can obtain the line
intensity as ${\cal F}L_{\rm line}^{\rm max}$. Inversely, if we
obtain ${\cal F}$ by comparing observed line luminosity and
theoretical $L_{\rm line}^{\rm max}$, we can estimate $\nH$.
\label{fig:metal_dilution}}
\end{figure}

In Figure \ref{fig:metal_dilution}, we show ${\cal F}$ as a
function of $\nH$, where equation (\ref{eq:simplef}) is
applied. If we detect line emission intensity of high-redshift
galaxies, we can make a rough estimate of the gas density
by comparing the observed line intensity with
$L_{\rm line}^{\rm max}$. The line ratio between two types
of lines can also constrain the gas density.

\section{Typical Angular Size of Clustering}\label{app:angular}

The integrated light from galaxies should trace their spatial
fluctuation produced by gravitational clustering. Therefore,
if we measure the typical correlation angular scale of the
integrated light, we can test if the integrated light
really comes from galaxies. Here, we calculate the typical
correlation angular size of galaxies at a certain redshift.

The typical mass within a clustered region, $M_{\rm cl}$, is
estimated to be the mass scale corresponding to
$\sigma (M_{\rm cl})D(z)=1$, where $D(z)$ is the
growth factor of
perturbations ($D(0)=1$), and $\sigma (M)$ is the variance
of density field smoothed with mass scale $M$.
In this relation, $M_{\rm cl}$ corresponds to the mass scale
at the break of the Press-Schechter mass function.
We adopt $\sigma (M)$ given in
Appendix of Kitayama \& Suto (1996).
Once $M_{\rm cl}$ is obtained, we can estimate the
typical diameter of gravitationally clustered region,
$D_{\rm cl}$, as
\begin{eqnarray}
D_{\rm cl}=2\left(
\frac{3M_{\rm cl}}{4\pi\rho_{\rm c0}\Omega_{\rm M}(1+z)^3}
\right)^{1/3}\, ,
\end{eqnarray}
where $\rho_{\rm c0}$ is the critical density of the
universe at $z=0$ (i.e., $\rho_{\rm c0}\Omega_{\rm M}$ is the
mean mass density of the universe at $z=0$).

In this paper, we are particularly interested in $z\sim 6$.
Using $D(6)=0.21$, we obtain
$M_{\rm cl}=2\times 10^{10}~M_\odot$ for the cosmology assumed
in this paper. This corresponds to
the typical comoving size of 0.33 Mpc
(i.e., $D_{\rm cl}=0.05$ Mpc). This corresponds to an angular
scale of $10''$.


\begin{thebibliography}{99}
\bibitem[Abel et al.(1997)]{abel97} Abel, T., Anninos, P., Zhang, Y.,
    Norman, M. L. 1997, NewA, 2, 181
\bibitem[Abel et al.(2002)]{abel02} Abel, T., Bryan, G. L.,
    Norman, M. L. 2002, Science, 295, 93
\bibitem[Anders \& Grevesse(1989)]{anders89} Anders, E., \&
    Grevesse, N. 1989, Geochim.\ Cosmochim.\ Acta, 53, 197
\bibitem[Armus et al.(1998)]{armus98} Armus, L., Matthews, K.,
    Neugebauer, G., Soifer, B. T. 1998, \apj, 506, L89
\bibitem[Barkana(2002)]{barkana02} Barkana, R. 2002, NewA, 7, 85
\bibitem[Bernstein et al.(2002)]{bernstein02} Bernstein, R. A.,
    Freedman, W. L., Madore, B. F. 2002, \apj, 571, 56
\bibitem[Bromm et al.(2001)]{bromm01} Bromm, V., Coppi, P. S., \&
    Larson, R. B. 2002, ApJ, 564, 23
\bibitem[Buat, Deharveng, \& Donas(1989)]{buat89} Buat, V.,
    Deharveng, J. M., \& Donas, J. 1989, \aap, 223, 42
\bibitem[Cambr\'{e}si et al.(2001)]{canbresy01} Cambr\'{e}si, L.,
    Reach, W. T., Beichman, C. A., \& Jarrett, T. H. 2001, \apj,
    555. 563
\bibitem[Carroll et al.(1992)]{carroll92} Carroll, S. M.,
    Press, W. H., \& Turner, E. L. 1992, ARA\&A, 30, 499
\bibitem[Ciardi et al.(2000)]{ciardi00} Ciardi, B., Ferrara, A.,
    Governato, F., \& Jenkins, A. 2000, \mnras, 314, 611
\bibitem[Ciardi \& Loeb(2000)]{ciardiloeb00} Ciardi, B., \& Loeb, A.
    2000, \apj, 540, 687
\bibitem[Cole et al.(2000)]{cole00} Cole, S., Lacey, C. G.,
    Baugh, C. M., \& Frenk, C. S. 2000, \mnras, 319, 168
\bibitem[Cox(2000)]{cox00} Cox, A. N. 2000, Allen's Astrophysical
    Quantities (4th ed.; New York: Springer)
\bibitem[Dale et al.(2001)]{dale01} Dale, D. A., Helou, G.,
    Neugebauer, G., Soifer, B. T., Frayer, D. T., \& Condon, J. J.
    2001, \aj, 122, 1736
\bibitem[Devriendt \& Guiderdoni(2000)]{devriendt00}
    Devriendt, J. E. G., \& Guiderdoni, B. 2000, \aap, 363, 851
\bibitem[Draine \& Bertoldi(1996)]{draine96} Draine, B. T., \&
    Bertoldi, F. 1996, \apj, 468, 269
\bibitem[Draine \& Lee(1984)]{draine84} Draine, B. T., \&  Lee, H. M.
    1984, \apj, 285, 89
\bibitem[Dwek et al.(1983)]{dwek83} Dwek, E., et al.\ 1983, \apj, 274,
    168
\bibitem[Dwek et al.(1998)]{dwek98} Dwek, E., et al.\ 1998, \apj, 508,
    106
\bibitem[Edmunds(2001)]{edmunds01} Edmunds, M. G. 2001, \mnras, 328,
    223
\bibitem[Elbaz et al.(2002)]{elbaz02} Elbaz, D., Cesarsky, C. J.,
    Chanial, P., Aussel, H., Franceschini, A., Fadda, D., \&
    Chary, R. R. 2002, \aap, 384, 848
\bibitem[Ferguson et al.(2002)]{ferguson02} Ferguson, H. C.,
    Dickinson, M., \& Papovich, C. 2002, \apj, 569, L65
\bibitem[Ferrara(1998)]{ferrara98} Ferrara, A. 1998, \apj, 499, L17
\bibitem[Ferrara, Pettini, \& Shchekinov(2000)]{ferrara00} Ferrara, A.,
    Pettini, M., \& Shchekinov, Y. 2000, \mnras, 319, 539
\bibitem[Fixsen et al.(1998)]{fixsen98} Fixsen, D. J., Dwek, E.,
    Mather, J. C., Bennett, C. L., \& Shafer, R. A. 1998, \apj, 508,
    123
\bibitem[Frenklach \& Feigelson(1997)]{frenklach97} Frenklach, M.,
    \& Feigelson, E. 1997, in ASP Conf.\ Ser.\ 122, From Stardust
    to Planetesimals, ed.\ Y. J. Pendleton \& A. G. G. M. Tielens
    (ASP: San Francisco), 107
\bibitem[Galli \& Palla(1998)]{galli98} Galli, D., \& Palla, F. 1998,
    \aap, 335, 403
\bibitem[Gardner \& Satyapal(2000)]{gardner00} Gardner, J. P., \&
    Satyapal, S. 2000, \aj, 119, 2589
\bibitem[Ge et al.(2001)]{ge01} Ge, J., Bechtold, J., \& Kulkarni,
    P. 2001, \apj, 547, L1
\bibitem[Gehrz(1989)]{gehrz89} Gehrz, R. D. 1989, in IAU Symp.\
    135, Interstellar Dust, ed.\ L. J.
    Allamandola \& A. G. G. M. Tielens (Dordrecht: Kluwer), 445
\bibitem[Gispert et al.(2000)]{gispert00} Gispert, R., Lagache, G.,
    \& Puget, J. L. 2000, \aap, 360, 1
\bibitem[Gorjian et al.(2000)]{gorjian00} Gorjian, V., Wright, E. L.,
    \& Chary, R. R. 2000, \apj, 536, 550
\bibitem[Granato et al.(2000)]{granato00} Granato, G. L., Lacey, C. G.,
    Silva, L., Bressan, A., Baugh, C. M., Cole, S., \& Frenk, C. S.
    2000, \apj, 542, 710
\bibitem[Haiman et al.(1996)]{haiman96} Haiman, Z., Rees, M. J., \&
    Loeb, A. 1996, \apj, 467, 522
\bibitem[Hauser et al.(1998)]{hauser98} Hauser, M. G., et al.\ 1998,
    \apj, 508, 25
\bibitem[Hauser \& Dwek(2001)]{hauser01} Hauser, M. G., \& Dwek, E.
    2001, ARA\&A, 39, 249
\bibitem[Hirashita, Hunt, \& Ferrara(2002a)]{hirashita02a} Hirashita, H.,
    Hunt, L. K., \& Ferrara, A. 2002a, \mnras, 330, L19
\bibitem[Hirashita, Tajiri, \& Kamaya(2002b)]{hirashita02b}
    Hirashita, H., Tajiri, Y. Y., \& Kamaya, H. 2002b, \aap, 388, 439
\bibitem[Hollenbach \& McKee(1979)]{hollenbach79} Hollenbach, D. J.,
    \& McKee, C. F. 1979, ApJS, 41, 555
\bibitem[Hu et al.(2002)]{hu02} Hu, E. M., Cowie, L. L.,
    McMahon, R. J., Capak, P., Iwamuro, F., Kneib, J.-P., Maihara, T.,
    \& Motohara, K. 2002, ApJL, 568, L75
\bibitem[Hutchings et al.(2002)]{hutchings02} Hutchings, R. M.,
    Santoro, F., Thomas, P. A., \& Couchman, H. M. P. 2002, \mnras,
    330, 927
\bibitem[Ikeuchi(1988)]{ikeuchi88} Ikeuchi, S. 1988, Fundam.\ Cosmic
    Phys., 12, 255
\bibitem[Inoue, Hirashita, \& Kamaya(2000b)]{inoue00b} Inoue, A. K.,
    Hirashita, H., \& Kamaya, H. 2000, \aj, 120, 2415
\bibitem[Jones, Tielens, \& Hollenbach(1996)]{jones96} Jones, A. P.,
    Tielens, A. G. G. M., \& Hollenbach, D. J. 1996, \apj, 469, 740
\bibitem[Kamaya \& Silk(2002)]{kamaya02} Kamaya, H., \& Silk, J. 2002,
    \mnras, 332, 251
\bibitem[Katz et al.(1999)]{katz99} Katz, N. Furman, I., Biham, O.,
    Pirronello, V., \& Vidali, G. 1999, \apj, 522, 305
\bibitem[Kauffmann \& Charlot(1998)]{kauffmann98} Kauffmann, G., \&
    Charlot, S. 1998, \mnras, 294, 705
\bibitem[Kennicutt(1998)]{kennicutt98} Kennicutt, R. C., Jr.\ 1998,
    \apj, 498, 541
\bibitem[Kitayama \& Ikeuchi(2000)]{kitayama00} Kitayama, T., \&
    Ikeuchi, S. 2000, \apj, 529, 615
\bibitem[Kitayama et al.(2001)]{kitayama01} Kitayama, T.,
    Susa, H., Umemura, M., \& Ikeuchi, S. 2001, \mnras, 326, 1353
\bibitem[Kitayama \& Suto(1996)]{kitayama96} Kitayama, T., \&
    Suto, Y. 1996, \apj, 469, 480
\bibitem[Kozasa, Hasegawa, \& Nomoto(1991)]{kozasa91} Kozasa, T.,
    Hasegawa, W., \& Nomoto, K. 1991, \aap, 249, 474
\bibitem[Lagache et al.(2000)]{lagache00} Lagache, G.,
    Haffner, L. M., Reynolds, R. J., \& Tufte, S. L. 2000, \aap,
    354, 247
\bibitem[Levshakov et al.(2002)]{levshakov02} Levshakov, S. A.,
    Dessauges-Zavadsky, M.,
    D'Odorico, S., \& Molaro, P. 2002, \apj, 565, 696
\bibitem[Lisenfeld \& Ferrara(1998)]{lisenfeld98} Lisenfeld, U.,
    \& Ferrara, A. 1998, \apj, 496, 145
\bibitem[Liszt(2002)]{liszt02} Liszt, H. 2002, \aap, 389, 393
\bibitem[Madau et al.(1996)]{madau96} Madau, P., Ferguson, H. C.,
    Dickinson, M., Giavalisco, M., Steidel, C. C., \& Fruchter, A.
    1996, \mnras, 283, 1388
\bibitem[Madau, Pozzetti, \& Dickinson(1998)]{madau98} Madau, P.,
    Pozzetti, L., Dickinson, M. 1998, \apj, 498, 106
\bibitem[Madau, Ferrara, \& Rees(2001)]{madau01} Madau, P.,
    Ferrara, A., \& Rees, M. J. 2001, \apj, 555, 92
\bibitem[Mather \& Stockman(2000)]{mather00} Mather, J. C., \&
    Stockman, H. S. 2000, The Institute of Space and
    Astronautical Science Report SP No.\ 14, Mid- and
    Far-Infrared Astronomy and Future Space Missions, ed.\
    T. Matsumoto \& H. Shibai (Sagamihara: ISAS), 203
\bibitem[Malkan \& Stecker(2001)]{malkan01} Malkan, M. A., \&
    Stecker, F. W. 2001, \apj, 555, 641
\bibitem[Matsuda, Sato, \& Takeda(1969)]{matsuda69} Matsuda, T.,
    Sato, H., \& Takeda, H. 1969, Prog.\ Theor.\ Phys., 42, 219
\bibitem[McKee(1989)]{mckee89} McKee, C. F. 1989, in IAU Symp.\
    135, Interstellar Dust, ed. L. J.
    Allamandola \& A. G. G. M. Tielens (Dordrecht: Kluwer), 431
\bibitem[McKee \& Ostriker(1977)]{mckee77} McKee, C. F., \&
    Ostriker, J. P. 1977, \apj, 218, 148
\bibitem[Mo \& White(2002)]{mo02} Mo, H. J., \& White, S. D. M.
    2002, \mnras, submitted (astro-ph/0202393)
\bibitem[Moseley et al.(1989)]{moseley89} Moseley, S. H., Dwek, E.,
    Glaccum, W., Graham, J. R., Loewenstein, R. F., \&
    Silverberg, R. F. 1989, Nature, 340, 697
\bibitem[Nakamura \& Umemura(2002)]{nakamura02} Nakamura, F., \&
    Umemura, M. 2002, \apj, 569, 549
\bibitem[Nagashima et al.(2001)]{nagashima01}
    Nagashima, M., Totani, T., Gouda, N., \& Yoshii, Y. 2001, \apj,
    557, 505
\bibitem[Nishi \& Susa(1999)]{nishi99} Nishi, R., \& Susa, H. 1999,
    \apj, 523, L103
\bibitem[Norman \& Spaans(1996)]{norman96} Norman, C. A., \&
    Spaans, M. 1996, \apj, 480, 145
\bibitem[Oh et al.(2002)]{oh02} Oh, S. P. 2002, \mnras, submitted
\bibitem[Omukai(2000)]{omukai00} Omukai, K. 2000, \apj, 534, 809
\bibitem[Omukai \& Nishi(1998)]{omukai98} Omukai, K., \& Nishi, R.
    1998, ApJ, 508, 141
\bibitem[Ouchi et al.(2002)]{ouchi02} Ouchi, M., et al.\ 2002,
    in RevMexAA Conf.\ Ser., Galaxy Evolution: Theory and
    Observations, in press
\bibitem[Partridge \& Peebles(1967)]{partridge67} Partridge, R. B.,
    \& Peebles, P. J. E. 1967, \apj, 148, 377
\bibitem[Pearson(2001)]{pearson01} Pearson, C. P. 2001, \mnras,
    325, 1511
\bibitem[Peebles(1980)]{peebles80} Peebles, P. J. E. 1980, The
    Large Scale Structure of the Universe (Peinceton: Princeton
    University Press)
\bibitem[Pei, Fall, \& Hauser(1999)]{pei99} Pei, Y. C., Fall, S. M.,
    \& Hauser, M. G. 1999, \apj, 522, 604
\bibitem[Petitjean et al.(2000)]{petitjean00} Petitjean, P.,
    Srianand, R., \& Ledoux, C. 2000, \aap, 364, L26
\bibitem[Pettini(2001)]{pettini01} Pettini, M. 2001, in 17th IAP
    Astrophysics Colloquium, Gaseous Matter in Galaxies and
    Intergalactic Spece, ed.\ R. Felet et al.\ (Frontier Group:
    Paris), 315
\bibitem[Press \& Schechter(1974)]{press74} Press, W. H., \& Schechter,
    P. 1974, \apj, 187, 425
\bibitem[Puget et al.(1996)]{puget96} Puget, J.-L., Abergel, A.,
    Bernard, J.-P., Boulanger, F., Burton, W. B., D\'{e}sert, F.-X.,
    \& Hartmann, D. 1996, \aap, 308, L5
\bibitem[Rana \& Wilkinson(1986)]{rana86} Rana, N. C., \& Wilkinson,
    D. A. 1986, MNRAS, 218, 497
\bibitem[Ripamonti et al.(2002)]{ripamonti02} Ripamonti, E., Haardt,
    F., Ferrara, A., \& Colpi, M. 2002, \mnras, in press
\bibitem[Roche \& Eales(1999)]{roche99} Roche, N., \& Eales, S. A.
    1999, \mnras, 307, 111
\bibitem[Salvaterra \& Ferrara(2002)]{salvaterra02} Salvaterra, R.,
    \& Ferrara, A. 2002, in preparation
\bibitem[Sanders \& Mirabel(1996)]{sanders96} Sanders, D. B., \&
    Mirabel, I. F. 1996, ARA\&A, 34, 749
\newpage
\bibitem[Santos et al.(2002)]{santos02} Santos, M. R., Bromm, V., \&
    Kamionkowski, M. 2002, \mnras, submitted (astro-ph/0111467)
\bibitem[Schaerer(2002)]{schaerer02} Schaerer, D. 2002, \aap, 382,
    28
\bibitem[Smail et al.(1998)]{smail98} Smail, I., Ivison, R. J., Blain,
    A. W., \& Kneib, J.-P. 1998, \apj, 507, L21
\bibitem[Shanks et al.(2001)]{shanks01} Shanks, T., Metcalfe, N.,
    Fong, D., McCracken, H., Campos, A., \& Thompson, D. 2001,
    in IAU Sump.\ 204, The Extragalactic Infrared Background and
    its Cosmological Implications, ed.\ M. Harwit \& M. G. Hauser
    (ASP: San Francisco), 347
\bibitem[Shapiro \& Kang(1987)]{shapiro87} Shapiro, P., \& Kang, H.,
    \apj, 318, 32
\bibitem[Soifer et al.(1998)]{soifer98} Soifer, B. T., Neugebauer, G.,
    Franx, M., Matthews, K., Illingworth, G. D. 1998, \apj, 501, L171
\bibitem[Somerville \& Primack(1999)]{somerville99} Somerville, S. R.,
    \& Primack, J. R. 1999, \mnras, 310, 1087
\bibitem[Spitzer(1978)]{spitzer78} Spitzer, L., Jr.\ 1978, Physical
    Processes in the Interstellar Medium (New York: Wiley)
\bibitem[Steidel et al.(1999)]{steidel99} Steidel, C. C., Adelberger,
    K. L., Giavalisco, M., Dickinson, M., \& Pettini, M. 1999, \apj,
    519, 1
\bibitem[Steidel et al.(1996)]{steidel96} Steidel, C. C.,
    Giavalisco, M., Pettini, M., Dickinson, M., \&
    Adelberger, K. L. 1996, \apj, 462, L17
\bibitem[Suginihara et al.(1999)]{suginohara99} Suginohara, M.,
    Suginohara, T., \& Spergel, D. N. 1999, \apj, 512, 547
\bibitem[Takeuchi et al.(2001a)]{t2_01a} Takeuchi, T. T., Ishii, T.
    T., Hirashita, H., Yoshikawa, K., Matsuhara, H., Kawara, K.,
    \& Okuda, H. 2001a, \pasj, 53, 37
\bibitem[Takeuchi et al.(2001b)]{t2_01b} Takeuchi, T. T., Kawabe, R.,
    Kohno, K., Nakanishi, K., Ishii, T. T., Hirashita, H., \&
    Yoshikawa, K. 2001b, PASP, 113, 586
\bibitem[Tan et al.(1999)]{tan99} Tan, J. C., Silk, J., \& Balland,
    C. 1999, \apj, 522, 579
\bibitem[Tegmark et al.(1997)]{tegmark97} Tegmark, M., Silk, J.,
    Rees, M. J., Blanchard, A., Abel, T., \& Palla, F. 1997, \apj,
    474, 1
\bibitem[Tielens \& Hollenbach(1985)]{tielens85} Tielens, A. G. G. M.,
    \& Hollenbach, D. 1985, \apj, 291, 722
\bibitem[Tinsley(1980)]{tinsley80} Tinsley, B. M. 1980, Fundam.\
    Cosmic Phys., 5, 287
\bibitem[Tinsley \& Danly(1980)]{tinsley_danly80} Tinsley, B. M., \&
    Danly, L. 1980, \apj, 242, 435
\bibitem[Todini \& Ferrara(2001)]{todini01} Todini, P., \&
    Ferrara, A. 2001, \mnras, 325, 726
\bibitem[Tosi \& Diaz(1990)]{tosi90} Tosi, M., \& Diaz, A. I. 1990,
    \mnras, 246, 616
\bibitem[Totani \& Takeuchi(2002)]{totani02} Totani, T., \&
    Takeuchi, T. T. 2002, \apj, 570, 470
\bibitem[Totani et al.(2001)]{totani01} Totani, T., Yoshii, Y.,
    Iwamuro, F., Maihara, T., \& Motohara, K. 2001, \apj, 550,
    L137
\bibitem[Walter et al.(2002)]{walter02} Walter, F., Weiss, A.,
    Martin, C., \& Scoville, N. 2002, \aj, 123, 225
\bibitem[White \& Frenk(1991)]{white91} White, S. D. M., \&
    Frenk, C. S. 1991, \apj, 379, 52
\bibitem[Wilson et al.(2000)]{wilson00} Wilson, C. D., Scoville, N.,
    Madden, S. C., \& Charmandaris, V. 2000, \apj, 542, 120
\bibitem[Woosley \& Weaver(1995)]{woosley95} Woosley, S. E., \&
    Weaver, T. A. 1995, \apjs, 101, 181
\bibitem[Wright(2001)]{wright01} Wright, E. L. 2001, \apj, 553, 538
\bibitem[Wright \& Reese(2000)]{wright00} Wright, E. L., \&
    Reese, E. D. 2000, \apj, 545, 53
\bibitem[Xu et al.(2001)]{xu01} Xu, C., Lonsdale, C. J., Shupe, D. L.,
    O'Linger, J., \& Masci, F. 2001, \apj, 562, 179
\end{thebibliography}
\end{document}